\documentclass[letter,12pt]{article}
\usepackage[subpreambles=false]{standalone} % Package for compilation of independent files

\usepackage[margin=3cm]{geometry}

\usepackage[utf8]{inputenc}
\usepackage[T1]{fontenc}

\usepackage{authblk}  % to add author affiliations

\usepackage[colorlinks]{hyperref}
\hypersetup{
  colorlinks,
  citecolor=red,
  linkcolor=blue,
  urlcolor=blue
}

\usepackage{sectsty,textcase}
\sectionfont{\fontsize{15}{15}\selectfont\MakeTextUppercase}
\subsectionfont{\fontsize{13}{15}\selectfont\MakeTextUppercase}
\subsubsectionfont{\fontsize{13}{15}\selectfont\MakeTextUppercase}

\usepackage[nolist]{acronym} % to add accronyms
\usepackage{url} % to add url hyperlink
\usepackage[title]{appendix} % to print "Appendix X" format
\usepackage{lineno} % to line numbering
\usepackage{amsmath,amssymb,amsfonts} % to add special symbols
\usepackage[misc,geometry]{ifsym} % to add special symbols
\usepackage{setspace} % to increase interline spacing
\usepackage{enumitem} % to remove the spacing in ITEMIZE
\usepackage{siunitx} % to format SI units
\usepackage[version=4]{mhchem} % to enable chemical representation
\usepackage{multirow} % to allow multirows within tables
\usepackage[flushleft]{threeparttable} % to allow footnotes right below tables
\usepackage{subcaption} % to enable subcaptions
\usepackage{cite} % to arrange citations
\usepackage[document]{ragged2e} % to avoid text justification
\usepackage[labelfont=bf,singlelinecheck=false,justification=justified]{caption} % to modify caption style
\usepackage[nomarkers,nolists,tablesfirst]{endfloat} % to put all figures at the end of the document  %%%% ACTIVATE TO SEND FIGURES TO THE END
\usepackage{graphicx} % to include inset figures
\usepackage{tcolorbox} % to include color text boxes
\usepackage{xcolor} % to have a bigger color palette
\definecolor{textboxyellow}{RGB}{255, 255, 204}

\providecommand{\keywords}[1]
{
  \small	
  \textbf{\textit{Keywords---}} #1
}

\title{\vspace{-2cm}Carbon fiber damage evolution under flame attack and the role of impurities}

\author[a,b,*]{Pablo Chávez-Gómez}
\author[a]{Tanja Pelzmann}
\author[c]{Darren R. Hall}
\author[c]{Cornelia Chilian}
\author[b]{Louis Laberge Lebel}
\author[a,*]{Étienne Robert}
\date{}

\affil[a]{{\footnotesize Multiphase and Reactive Flows Laboratory, Department of Mechanical Engineering}}
\affil[b]{{\footnotesize Advanced Composite and Fiber Structures Laboratory (ACFSLab) / Research Center for High Performance Polymer and Composite Systems (CREPEC), Department of Mechanical Engineering}}
\affil[c]{{\footnotesize SLOWPOKE Neutron Activation Analysis Laboratory, Department of Engineering Physics}}
\affil[ ]{{Polytechnique Montréal, Montréal, QC, H3T 1J4, Canada\vspace{0.25cm}}}
\affil[*]{{\footnotesize Corresponding authors: \href{mailto:pablo.chavez@polymtl.ca}{pablo.chavez@polymtl.ca} / \href{mailto:etienne.robert@polymtl.ca}{etienne.robert@polymtl.ca}}}

\begin{document}

\maketitle

\vspace{-1cm}
\begin{abstract}

\justifying
\Acp{CF} are prone to extensive oxidation under fire attack, for instance, in an aircraft fire scenario. This work addresses the damage mechanisms observed on \ac{PAN}-based \acp{CF} with different microstructure exposed to open flames. A fixed-point technique was developed to follow up individual \acp{CF} by means of time-controlled insertion into premixed methane/air flames, followed by \ac{SEM} and \ac{EDS} analyses. Besides diameter reduction, three localized damage mechanisms were discerned in presence of impurities, which were quantified by \ac{NAA}. Severe pitting was ascribed to catalytic oxidation mainly caused by alkali and alkaline earth metals. After an initial period where catalytic reactions between impurities and the carbon surface dominate, the flame stoichiometry governed the \ac{CF} gasification process, with lean flames being much more aggressive than rich ones. A second mechanism, channelling, was caused by mobile metallic impurities. Some impurities showed an opposite effect, lowering reactivity and thus preventing further catalysis. Amorphous damage with a skin-peeling effect is believed to be the result of localized impurities at high concentrations and microstructural variations. Hindrance or synergistic effects between impurities are discussed. Finally, apparent axial pit growth rates were determined and compared with other carbonaceous materials, revealing a strong influence of impurities and the flame reactive atmosphere on \ac{CF} oxidation.
\end{abstract}


%\section*{Nomenclature}
\begin{acronym}[XXXXXXXXXX]\itemsep=1pt

\acro{AFM}{atomic force microscopy}
\acro{ASA}{active surface area}
\acro{CASA}{contact active surface area}
\acro{CF}{carbon fiber}
\acro{CI}{confidence interval}
\acro{CFRP}{carbon fiber reinforced polymer}
\acro{CMC}{ceramic matrix composite}
\acro{CNT}{carbon nanotube}
\acro{DSC}{differential scanning calorimetry}
\acro{EDS}{energy-dispersive X-ray spectroscopy}
\acro{Ea}[$E_a$]{activation energy}
\acro{ESEM}{environmental scanning electron microscopy}
\acro{ETEM}{environmental transmission electron microscopy}
\acro{FAA}{Federal Aviation Administration}
\acro{FFB}{flat flame burner}
\acro{HOPG}{highly oriented pyrolytic graphite}
\acro{HTSEM}[HT-SEM]{high temperature environmental scanning electron microscopy}
\acro{HTT}{heat treatment temperature}
\acro{HUR}{heat-up rate}
\acro{ISO}{International Organization for Standardization}
\acro{NAA}{neutron activation analysis}
\acro{PAN}{polyacrylonitrile}
\acro{PMC}{polymer matrix composite}
\acro{Qdot}[$\dot{Q}$]{heat flux}
\acro{SEM}{scanning electron microscopy}
\acro{STM}{scanning tunneling microscopy}
\acro{STEM}{scanning transmission electron microscopy}
\acro{TEM}{transmission electron microscopy}
\acro{SLOWPOKE}[SLOWPOKE-2]{Safe LOW-POwer Kritical Experiment}
\acro{SS}{stainless steel}
\acro{STA}{simultaneous thermal analyses}
\acro{T}[$T$]{temperature}
\acro{TTa}[$T_{Ta}$]{Tammann temperature}
\acro{Tm}[$T_m$]{melting point}
\acro{Tmax}[$T_{max}$]{temperature corresponding to the maximum oxidation rate}
\acro{TEM}{transmission electron microscopy}
\acro{TGA}{thermogravimetric analysis}
\acro{TPS}{thermal protection systems}
\acro{TTF}{time-to-failure}
\end{acronym}

%\linenumbers
\begin{spacing}{2.0}

\section{Introduction} \label{sec:Introduction}

In-flight or post-crash aircraft fires pose a threat to passenger safety, either from a smoke and toxicity standpoint, or repercussions on the vehicle structural integrity \cite{DOT-FAA-AR-TN11-8improvements}. This translates into challenging design constraints for modern aircraft, since they extensively rely on inherently-flammable \ac{CFRP} composites \cite{Mouritz2006book}. The intricate thermal, physical and chemical processes involved in \ac{CFRP} combustion \cite{Langot2021multiphysics, Langot2021reactionscheme} make the fate of \acp{CF} exposed to fire difficult to predict. Moreover, desirable outcomes are sometimes conflicting. On one hand, full gasification might be sought \cite{Hull1981fibergasification, Ganjei1981catalyzedoxidation} to avoid the release of \ac{CF} fragments from burning composites, resulting in health and electrical hazards \cite{Sussholz1980fiberrelease, Gandhi1999potential, Mouritz2009toxicity}. On the other hand, oxidative resistance is required to prevent burn-through if the component is structural or serves as a firewall \cite{DOT-FAA-AR-TN11-8improvements}. However, the conditions under which this resistance is assessed varies, with samples typically exposed to the atmospheres created by non-premixed or partially premixed turbulent flames.

\Acp{CF} readily gasify at high temperatures when exposed to reactive or oxidizing environments. Like any other carbonaceous material \cite{WalkerJr1959carbonoxygenreaction}, the fiber reactivity is mainly driven by its \ac{ASA} \cite{Ismail1991reactivity, Hippo1989activesites}, which is defined by structural order and available functional groups \cite{Figueiredo2010surfacechemistry}. \ac{CF} oxidation takes place through preferential etching starting at defects (e.g. vacancies, Stone-Wales type, interstitials, adatoms \cite{Terrones2010graphene}) and crystal edges (armchair \& zig-zag) \cite{Ismail1987activesurface, Hippo1989activesites, Lee1999defectinduced, Hahn1999mechanistic, Hahn2005kinetic}. Such features mainly arise during the fiber manufacturing process \cite{Morgan2005bookCFs, Newcomb2016carbonfibers}. The heterogeneous \ac{CF} microstructure involving turbostratic and amorphous regions \cite{Perret1970microstructure, Barnet1973oxygenplasma, Guigon1984highstrength, Guigon1984highmodulus, Morgan2005bookCFs} results in intricate oxidation processes \cite{Shi2021oxidation}. Moreover, fibers from the same precursor, e.g. \ac{PAN}, have different structures depending on the \ac{HTT}. High strength/low modulus \acp{CF} show a core-sheath structure with small crystallites in the skin zone and regions of amorphous carbon in the core \cite{Guigon1984highstrength}, while high modulus \acp{CF} have a less disordered structure \cite{Guigon1984highmodulus} as aresult of higher \ac{HTT}. 

Fiber gasification is influenced by several factors, namely the composition of the atmosphere, temperature and pressure \cite{Ismail1991reactivity, Halbig2008oxidation, Kim2011commercial, Govorov2016kinetics, Figueiredo2010surfacechemistry, ChavezGomez2022flamechemistry}. For instance, with molecular oxygen (\ce{O2}), oxidation takes place by \ce{O2} chemisorption and subsequent carbon monoxide and dioxide (\ce{CO}/\ce{CO2}) desorption. Other mechanisms can come into play in the absence of \ce{O2}, such as the Boudouard reaction in the presence of \ce{CO2} \cite{WalkerJr1959carbonoxygenreaction}. These mechanisms enlarge pores, create more active sites and new defects thus increasing the \ac{ASA}, enhancing the oxidation process and degrading the \ac{CF} mechanical properties \cite{Tong2011kineticstensile, Feih2012tensilefire, Bertran2013oxidation, Kachold2016preheating, Vautard2014defect}, with analogous processes observed with other carbonaceous materials \cite{Pickup1986fracture}. These phenomena can be accelerated by the presence of impurities in the carbon structure, which act locally or with a certain mobility \cite{Baker1979insitucatlayst, Baker1982TammannTemp, Baker1986catalyst, SousaLobo2016kinetics}. As with other carbonaceous materials, alkali and alkaline earth metals, along with their carbonate and acetate compounds, are known to catalyze the \ac{CF} oxidation process \cite{McMahon1978oxidativeresistance, Gibbs1979stability, Ganjei1981catalyzedoxidation, Hull1981fibergasification, Eckstein1981oxidation, Scola1988oxidation, Ismail1991reactivity}. Other elements can mitigate catalytic effects, for instance boron doping results in active site blockage and crystallite size increase \cite{Jones1991boroninfluence, Wu2005boroninhibition}, while some halogens, sulphur (\ce{S}) and phosphorus (\ce{P}) can poison impurities with catalytic effects \cite{Otto1979sulfureffect, Wu2006phosphorusinhibition}.

\ac{CF} gasification invariably leads to diameter reduction \cite{Yin1994oxidation, Feih2012tensilefire} and internal porosity may develop in a process analogous to carbon activation, i.e., controlled pore and \ac{ASA} enlargement \cite{Ko1992activation, Fuertes1996gasification}. However, locally accelerated oxidation may take place in the presence of impurities and large structural defects. This process, known as pitting, has been reported on \acp{CF} under liquid or gaseous oxidative conditions, such as electrolyte \cite{Marshall1991topography} and acid treatments \cite{Hoffman1992scanning, PittmanJr1997nitricacid}, \ac{TGA} \cite{Ismail1991reactivity, Serp1997vaporgrownCF, Cho1997protection, Halbig2008oxidation, Tzeng2010resistance, Bertran2013oxidation}, plasma etching \cite{Barnet1973oxygenplasma, Hoffman1992scanning}, as well as using tube furnaces \cite{Iacocca1993catalytic, Bertran2013oxidation}, flow tube reactors \cite{Panerai2013flow, Panerai2019experimentaloxidation}, air-filled ovens \cite{Tong2011kineticstensile, Ismail1991reactivity}, environmental electron microscopes \cite{Cochell2021nanoscaleoxidation} and other heated processing systems \cite{Kachold2016preheating}. Under open flame attack, exacerbated pitting was observed on \ac{CF} bundles after exposures of a few minutes to \ce{CH4}-based flames \cite{ChavezGomez2019fiberoxidation}. Pitting has also been observed on \ce{C}/\ce{C} composites after exposure to oxyacetylene flames \cite{Cho1993microscopicablation} as well as on burnt \ac{CFRP} composites. In flammability tests where samples are exposed to a radiative heat source, matrix decomposition and flaming combustion can expose \acp{CF} to the reactive atmosphere \cite{Eibl2017respirablefibers}. Samples have also been directly exposed to fuel pool fires to simulate post-crash flame conditions and investigate health hazards from \ac{CF} fragments released following oxidation-induced diameter reduction, pitting, and fibrillation \cite{Bell1979fiberburning, Sussholz1980fiberrelease}. However, the vast majority of the aforementioned studies were performed in environments that do not represent the conditions encountered in aircraft fire scenarios where continuous open flame attack is a threat. This is the case of post-crash conditions involving \ac{CFRP}-based fuselage skins or in-flight fires involving, for instance, powerplant firewalls or cargo liners. For certification purposes, standardized intermediate-scale tests \cite{AC20-135ch1, ISO2685, FAAFireHandbook} are necessary to determine, among other attributes, the burn-through resistance of the aforementioned structures. In such standardized tests, temperature, heat flux and fuel type are controlled, although little attention is given to flame chemistry. However, a previous work \cite{ChavezGomez2022flamechemistry} showed that, in addition to the mechanical loads, \ac{CF} failure is highly influenced by the flame stoichiometry. Thus, it is necessary to revisit the fire-induced carbon fiber damage processes.

Studies involving \ac{CF} exposure to controlled flame conditions are limited, specifically with regards to the characterization of the reactive atmosphere and the systematic investigations of \ac{CF} pitting. However, results obtained with other carbonaceous materials \cite{SousaLobo2016kinetics} can shed light on this phenomenon. For instance, pitting and impurity-induced channelling have been widely studied in graphite model materials, such as natural graphite \cite{Hughes1962topography, Thomas1964localizedoxidation, Baker1979insitucatlayst} and \ac{HOPG} \cite{Chang1990monolayer, Chang1991scanning, Chu1991gasificationSTM, Stevens1998kinetics, Lee1999defectinduced, Hahn1999mechanistic, Nicholson2003temperature, Nicholson2005etching, Hahn2005kinetic, Delehouze2011transition, Dobrik2013etching}, as well as in \acp{CNT} \cite{Shimada2004pointsonsetgasification} and graphene \cite{Johanek2016realtime}). The effect of the gaseous atmosphere composition on pitting has been assessed for air \cite{Lee1999defectinduced, Hahn1999mechanistic, Hahn2005kinetic}, pure oxygen (\ce{O2}) and mixtures \cite{Hughes1962topography, Thomas1964localizedoxidation, Chu1991gasificationSTM, Stevens1998kinetics, Lee1999defectinduced, Dobrik2013etching}, hydrogen \ce{H2} \cite{Tomita1974hydrogenation, Chu1991gasificationSTM}, and in presence of transient or atomic species (e.g. oxygen \ce{O}($^3$P) \cite{Nicholson2003temperature, Nicholson2005etching, Murray2018dynamics}, nitrogen (\ce{N}) and hydrogen (\ce{H}) \cite{McCarroll1970interaction}), revealing characteristic behaviors influenced by each material and testing condition. In these studies, \textit{in situ} or \textit{post hoc} observations have allowed the determination of pit growth rates. Numerical models dedicated to \ac{CF} pitting are being developed based on pit growth rates found in highly oxidative environments, for instance, in simulated atmospheric re-entry conditions \cite{Fu2022pitting}. Again, fire-induced oxidation and the role played by microstructure or impurities in pitting have not been investigated, and direct comparison between materials is not yet possible. This demonstrates the need for dedicated \ac{CF} pitting analyses, especially in aggressive and fire-representative environments. 

In view of these challenges, we investigated the fire-induced damage mechanisms of \ac{PAN}-based \acp{CF}, focusing on surface pitting and its growth rate. Other effects such as channelling, porosity development and amorphous erosion were also addressed. Considering three commercial \acp{CF}, we closely followed up these features after successive \ac{CF} insertions into premixed laminar methane (\ce{CH4})/air flames. Using a \acf{SEM}-based fixed-point technique, apparent pit growth rates were determined for standard and intermediate modulus \acp{CF}. \Ac{NAA} and \acf{EDS} were used to quantify and localize the presence of impurities. The pit growth values measured were then directly compared with values available in the literature for controlled conditions, i.e., atmosphere and graphite models.

%%%%%%%%%%%%%%%%%%%%%%%%%%%%%%%%%%%%%%%%%%%%%%%%%
% % % % % % % % % % % % % % % % % % % % % % % % %
%%%%%%%%%%%%%%%%%%%%%%%%%%%%%%%%%%%%%%%%%%%%%%%%%
\section{Experimental} \label{sec:Experimental}

\subsection{Materials} \label{sec:Experimental_Materials}
Three commercial \ac{PAN}-based \acp{CF} (Hextow\textsuperscript{\textregistered}, Hexcel) with different properties have been selected: AS4 \cite{Hexcel_AS4_TDS}, IM7 \cite{Hexcel_IM7_TDS} and HM63 \cite{Hexcel_HM63_TDS}, which correspond to standard, intermediate, and high modulus, respectively. The fiber bundles were taken from the same spools used in a previous work \cite{ChavezGomez2022flamechemistry}, enabling direct comparison with other flame exposure results.

\subsection{Fiber characterization} \label{sec:Experimental_Characterization}

Qualitative morphology analyses were carried out via \acf{SEM} on both virgin and burnt \acp{CF} using a high resolution field emission microscope (JSM7600F, JEOL). Acceleration voltage and current were set at \qty{6}{\kilo\volt} and \qty{226.4}{\micro\ampere}, respectively. Impurities on the surface of fibers, along with \ce{O}, \ce{N} and \ce{C} levels, were assessed with a built-in \acf{EDS} detector (X-Max\textsuperscript{N}, Oxford Instruments) using an energy range of \qty{10}{\kilo\electronvolt}. Measurements of fiber features were performed using the Fiji/ImageJ software \cite{Schneider2012imagej}.

The presence of potential oxidation catalysts was quantified through \acf{NAA} using a \ac{SLOWPOKE} nuclear reactor located at Polytechnique Montréal \cite{Townes1985slowpoke}. The specific details related to this method are described in detail elsewhere \cite{Abdollahineisiani2018NAA}. Two different samples for each type of \ac{CF} bundle were cut to a length of \qty{2.6}{\meter} and irradiated using a neutron flux of \qty{5e11}{\per\square\centi\meter\per\second}. One \ac{CF} sample was dedicated to short-lived radioisotopes, irradiated for \qty{10}{\minute} and counted upon a \qty{6}{\minute} decay. The second sample was intended for medium- and long-lived radioisotopes, irradiated for \qty{135}{\minute}, and a count was performed after 4- and 10-day decays, respectively.

%%%%%%%%%%%%%%%%%%%%%%%%%%%%%%%%%%%%%%%%%%%%%%%%%
% % % % % % % % % % % % % % % % % % % % % % % % %
%%%%%%%%%%%%%%%%%%%%%%%%%%%%%%%%%%%%%%%%%%%%%%%%%
\subsection{Controlled insertion}

\begin{figure}[htbp]
    \centering
    \includegraphics[width=0.5\columnwidth,height=0.90\textheight,keepaspectratio]{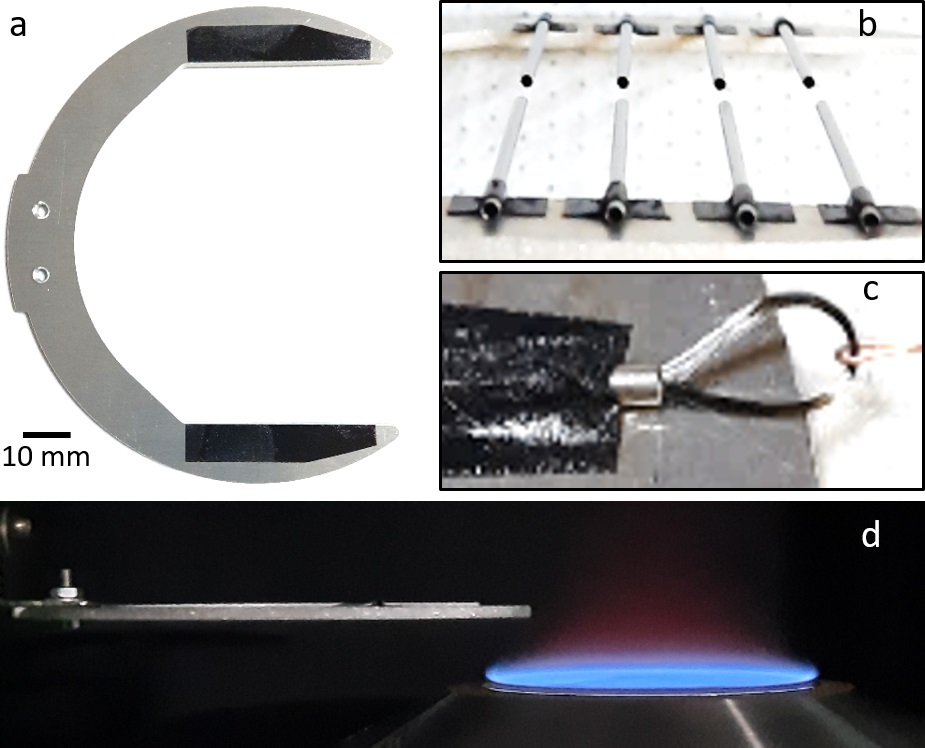}
    \caption{Top view of sample holder with carbon tape (a), stainless steel tubes (b), \ac{CF} bundle threading method (c), and side view (d) of the insertion method into the flame.}
    \label{fig:Sample_Holder_FFB}
\end{figure}

A fixed-point method was conceived, inspired by the \textit{post-hoc} observations of oven-based \acp{CNT} oxidation by Morishita \textit{et al.} \cite{Morishita1997CNTgasification, Morishita1999SEMpurificationCNT, Shimada2004pointsonsetgasification}. This allowed \ac{SEM}/\ac{EDS} analyses to be carried out with the same sample holder used to burn the \acp{CF}. A \ac{FFB} (standard model, Holthuis \& Associates) comprising a stainless steel body and a water-cooled bronze plate was used to obtain premixed fuel-lean, stoechiometric and rich \ce{CH4}/air flames, respectively with fuel-to-oxidizer equivalence ratios $\phi=\{0.7,1.0,1.2\}$ and $T\approx$ \qtylist[list-units = single]{1706;1790;1723}{\kelvin}. The flame properties, i.e., temperature and species concentrations, details on the \ac{FFB} configuration and gas measurements are described elsewhere \cite{Weigand2003, ChavezGomez2022flamechemistry}. Although flames found in real or simulated aircraft fires are typically turbulent and non-premixed, the use of this burner allowed to control the flame stoichiometry with minimal influence of other factors such as flame speed and turbulence, reducing fiber breakage and thus enabling the close follow-up of individual damage features. Fig. \ref{fig:Sample_Holder_FFB} shows the custom horseshoe-shaped aluminum 6061 sample holder designed to hold fiber samples during flame exposure and subsequent \ac{SEM} observations. Its inner contour was defined by the \ac{FFB}'s bronze outlet (OD = \qty{73.5}{\milli\meter}, plus \qty{1.75}{\milli\meter} clearance). The external contour allows self-aligned installation in the \ac{SEM} sample holder (SM-71090, JEOL). The horseshoe was fixed  using \qty{8}{\milli\meter} double-sided conductive carbon tape (No. 5028581, Fisher Scientific). To guide the fibers, AISI 304 \acf{SS} tubes (OD = \qty{1.2}{\milli\meter}, ID = \qty{1.0}{\milli\meter}, Unimed) were cut to a length of \qty{40}{\milli\meter} and bonded onto the horseshoe using an ethyl cyanoacrylate glue (Super Glue Gel Control\textsuperscript{\textregistered}, LePage) and carbon-taped to ensure electrical conductivity. Four pairs of tubes were installed \qty{8.5}{\milli\meter} from each other. The outermost pair of tubes was not used, and only served as visual reference. For fiber installation, a small \ac{CF} bundle was extracted from the spool and guided through a pair of aligned steel tubes with a Chromel wire, as shown in Fig. \ref{fig:Sample_Holder_FFB}c, using a needle threading-like technique. The bundles were fixed with the same carbon tape, followed by sequential cutting and removal from the central opening to decrease the amount of fibers that would be exposed to the flame. 

Fibers were intermittently exposed to the flame \qty{15}{\milli\meter} above the burner's porous surface. To maximize the accuracy of exposure time and height as well as to minimize the displacement time, the fibers were translated in and out the flame using a brushless DC-servomotor and a \qty{100}{mm} stroke rod (QUICKSHAFT\textsuperscript{\textregistered} LM1247-100-01, Faulhaber). The motion controller (MCLM 3006, Faulhaber) was set to yield the actuator's maximum speed, i.e., $\sim$\qty{3}{\meter\per\second}. The insertion time and the speed were verified using a high speed camera (FASTCAM Mini AX200, Photron) recording at 10,000 fps. Based on the sample holder tip position, the travel time (from start to stop) was estimated at $\sim$\qty{100}{ms}. The reported flame exposure values correspond to static residence time only.

%%%%%%%%%%%%%%%%%%%%%%%%
% RESULTS & DISCUSSION %
%%%%%%%%%%%%%%%%%%%%%%%%

\section{Results \& Discussion} 

\subsection{Virgin fibers} \label{sec:Results_Virgin_fibers}

%%%%%%%%%%%%%%%%%%%%%%%%%%%%%%%%%%%%%%%%%%%%%%%%%
% % % % % % % % % % % % % % % % % % % % % % % % %
%%%%%%%%%%%%%%%%%%%%%%%%%%%%%%%%%%%%%%%%%%%%%%%%%
\subsubsection{Morphology} \label{sec:Results_Virgin_Morphology}

\begin{figure}[htbp]
    \centering
    \includegraphics[width=\textwidth,height=0.90\textheight,keepaspectratio]{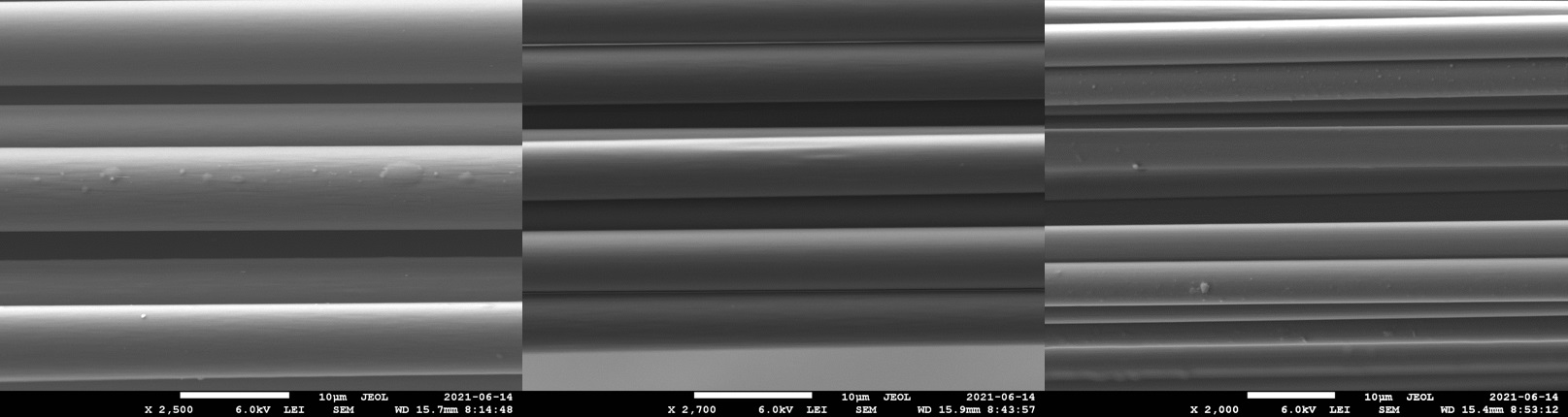}
    \caption{\ac{SEM} images of virgin AS4, IM7 and HM63 fibers (left, center \& right, respectively) as extracted from the fiber spool.}
    \label{fig:SEM_virgin_fibers}
\end{figure}

Fig. \ref{fig:SEM_virgin_fibers} shows \ac{SEM} images of the three different virgin fibers. Homogeneous surfaces with no bumps nor pits and constant diameters are prevalent. However, some features such as light striations running lengthwise can be seen on the unsized fibers (AS4 and IM7). Additional sub-micron features are observed on some of the AS4 and HM63 fibers. In both cases, they seem to be either particles fused with the fiber or foreign grain-like items, in the case of HM63 fibers. Moreover, some oval-shape bulges can be observed on the surface of some AS4 fibers. Several points were scanned along $\sim$\qty{25}{\micro\meter} of the fiber surface with \ac{EDS} and no significant elementary differences were observed when compared to smoother portions. This suggests the presence of buried impurities or surface defects with uniform chemical composition. \ac{EDS} measurements and elemental analysis are discussed below in \S \ref{sec:Results_EDS_Virgin_fibers} and \S \ref{sec:Results_Impurities}, respectively.

%%%%%%%%%%%%%%%%%%%%%%%%%%%%%%%%%%%%%%%%%%%%%%%%%
% % % % % % % % % % % % % % % % % % % % % % % % %
%%%%%%%%%%%%%%%%%%%%%%%%%%%%%%%%%%%%%%%%%%%%%%%%%
\subsubsection{Surface composition} \label{sec:Results_EDS_Virgin_fibers}

\begin{figure}[htbp]
    \centering
    \includegraphics[width=\textwidth,height=0.90\textheight,keepaspectratio]{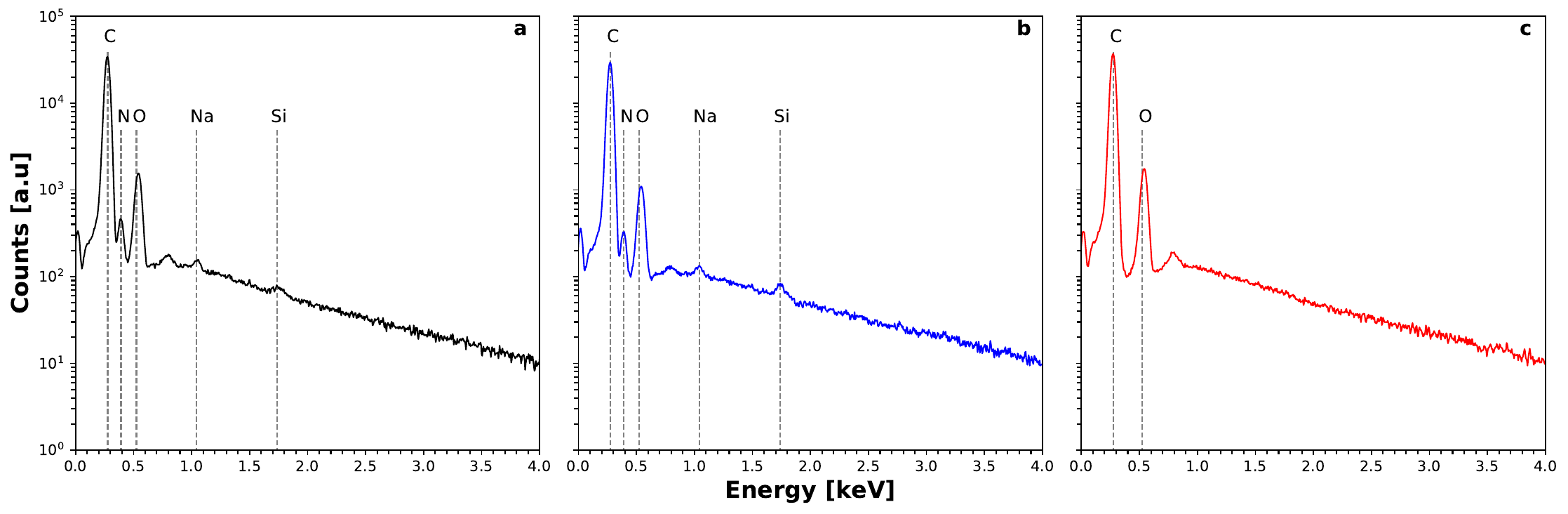}
    \caption{\ac{EDS} spectra of virgin AS4, IM7 and HM63 fibers (from left to right). All confirmed peaks correspond to the K$_{\alpha}$ values. The plots are shown in an unconventional semi-log scale for clarity of peaks above $\sim$\qty{1.0}{\kilo\eV}.}
    \label{fig:EDS_Virgin_fibers}
\end{figure}

Fig. \ref{fig:EDS_Virgin_fibers} shows the \ac{EDS} spectrum for each virgin fiber surface in a semi-log plot. The y-axis has been transformed to allow the identification of weak signals caused by minuscule impurities. Otherwise, in a linear scale, they are typically indiscernible in comparison with the dominating signatures of \ce{C}, \ce{N} and \ce{O}. The main peak describing the carbon content at \qty{0.277}{\kilo\eV} is self-evident. With respect to heteroatoms, i.e., \ce{O} and \ce{N}, two different situations are observed. The first involves AS4 (Fig. \ref{fig:EDS_Virgin_fibers}a) and IM7 (Fig. \ref{fig:EDS_Virgin_fibers}b) fibers, where \ce{O} and \ce{N} peaks are second and third in prominence at 0.525 and \qty{0.392}{\kilo\eV}, respectively. Their presence is expected in \ac{PAN}-based \acp{CF}, since \ce{N} atoms can be traced back to the precursor chains, whereas \ce{O} atoms are attributed to the stabilization step. Several surface functional groups are possible in the presence of \ce{N} and \ce{O} atoms \cite{Figueiredo2010surfacechemistry} within the basal planes or bound to the graphene layers' edges, providing sites with increased oxidative potential \cite{Boehm2012freeradicals}. Moreover, sodium (\ce{Na}) and silicon (\ce{Si}) impurities can be discerned from smaller peaks at 1.041 and \qty{1.739}{\kilo\eV}, respectively. The presence of \ce{Na} already suggests catalyzed oxidation upon exposure to flames, whereas \ce{Si} is not expected to react adversely with the fiber structure and actually has a deactivating effect on alkali and alkaline earth impurities \cite{Arnold2019catalysts}. Their concentration levels are discussed in detail at \S \ref{sec:Results_Impurities}. HM63 fibers show a  contrasting composition in Fig. \ref{fig:EDS_Virgin_fibers}c, since the \ce{N}, \ce{Na} and \ce{Si} peaks are absent, as opposed to AS4 or IM7 fibers. This might be explained by the higher \acp{HTT} that high modulus \ac{PAN}-based \acp{CF} undergo at the graphitization step, which helps to volatilize impurities and remove \ce{N} atoms from the original \ac{PAN} structure. 

%%%%%%%%%%%%%%%%%%%%%%%%%%%%%%%%%%%%%%%%%%%%%%%%%
% % % % % % % % % % % % % % % % % % % % % % % % %
%%%%%%%%%%%%%%%%%%%%%%%%%%%%%%%%%%%%%%%%%%%%%%%%%
\subsubsection{Impurity analysis} \label{sec:Results_Impurities}

\begin{figure}[htbp]
    \centering
    \includegraphics[width=0.5\columnwidth,height=0.90\textheight,keepaspectratio]{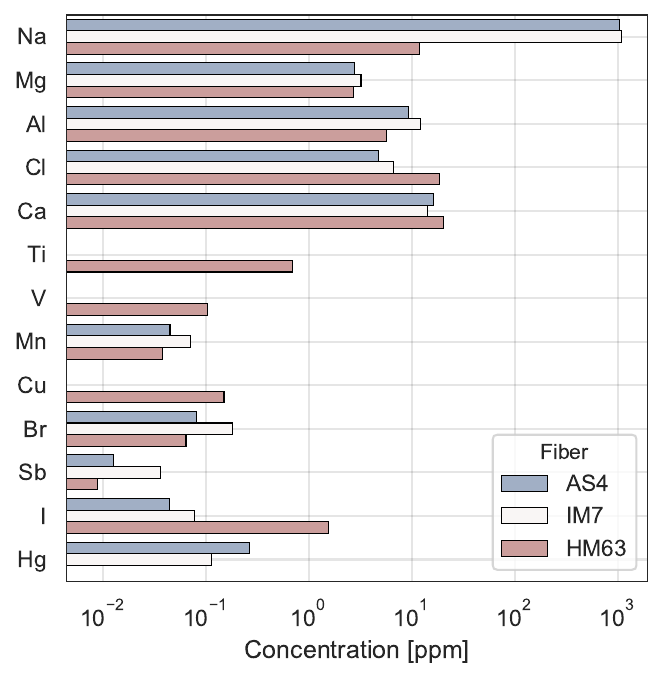}
    \caption{Impurity concentrations obtained via \ac{NAA} from standard (AS4), intermediate (IM7) and high modulus (HM63) fibers. Only fully confirmed elements are shown.}
    \label{fig:NAA_results}
\end{figure}

Fig. \ref{fig:NAA_results} shows the concentration of impurity elements (in ppm) detected via \ac{NAA} for the three fiber types. Only elements with well defined gamma-ray peaks are shown, namely \ce{Na}, calcium (\ce{Ca}), aluminum (\ce{Al}), chlorine (\ce{Cl}), magnesium (\ce{Mg}), iodine (\ce{I}), bromine (\ce{Br}), manganese (\ce{Mn}) and antimony (\ce{Sb}). The detection limits were not the same for all elements in all fibers, with certain species detected in only one or two fibers. Such is the case for AS4 and IM7 fibers, which yielded a quantifiable mercury \ce{Hg} content. Conversely, HM63 yielded titanium (\ce{Ti}), vanadium (\ce{V}), and copper (\ce{Cu}) atoms. The latter is noteworthy since some particles were visualized by \ac{SEM} and confirmed by \ac{EDS} to contain \ce{Cu}. This is further discussed in \S\ref{sec:Results_Burnt_Morphology}. The full \ac{NAA} data can be found in the supplementary material.

First, we address impurities that have a known catalytic effect. As an alkali metal, \ce{Na} is recognized as a very effective carbon oxidation promoter \cite{Amariglio1966combustioncatalytique, McMahon1978oxidativeresistance, Gibbs1979stability, Eckstein1981oxidation, Ganjei1981catalyzedoxidation, Ismail1991reactivity}. In our tests, it was found to be the most abundant impurity in AS4 and IM7 fibers at 1024$\pm$41 and 1079$\pm$43 ppm, respectively. In contrast, HM63 yielded a much lower concentration, 11.9$\pm$0.5 ppm. In addition to a higher crystallinity, such reduction by two orders of magnitude is believed to be a critical factor in HM63's increased oxidative resistance, as shown in tests under flame and mechanical load \cite{ChavezGomez2022flamechemistry}. The root causes of such marked differences in \ce{Na} concentration can be traced back to the processing steps of \ac{PAN}-based \acp{CF}, where the precursor can be dissolved using sodium thiocyanate (\ce{NaSCN}) in the polymerization step \cite{Newcomb2016carbonfibers}. Subsequently, it remains within the carbon structure through the oxidative stabilization and carbonization steps. Further graphitization helps to vaporize internal impurities \cite{Lieberman1972impurities,McMahon1978oxidativeresistance}, hence the lower \ce{Na} concentration in HM63 fibers. Additional \ce{Na}-based impurities may come from anodic oxidation which aim at improving fiber/matrix adhesion. This can be achieved with diverse electrolytes, namely sodium chloride (\ce{NaCl}) and sodium hypochlorite (\ce{NaOCl}) \cite{Morgan2005bookCFs}. These in turn can also explain the presence of \ce{Cl}, which is discussed below. Overall, \ce{Na} levels are in good agreement with those found in the literature for \ac{PAN}-based \acp{CF} \cite{McMahon1978oxidativeresistance, Eckstein1981oxidation, Scola1988oxidation, Ismail1991reactivity}. The presence of other alkalis could not be precisely determined. With regard to the confirmed alkaline earth metals, i.e., \ce{Ca} and \ce{Mg}, both elements are known to be effective catalysts. Different catalytic effects have been reported with related acetates and carbonates \cite{Ganjei1981catalyzedoxidation}. Their concentration levels are very similar in the three fiber types. \ce{Ca} concentrations were found at 16.2$\pm$2.2, 14.1$\pm$2.6. and 20.1$\pm$1.3 ppm, whereas \ce{Mg} yielded 2.74$\pm$0.75, 3.19$\pm$0.74 and 2.70$\pm$0.24 ppm for AS4, IM7 and HM63, respectively. Their effect is expected to be more pronounced in standard (AS4) and intermediate (IM7) \acp{CF}. These fibers have a higher amorphous carbon content and a more pronounced turbostratic structure than their high modulus (HM63) counterpart \cite{ChavezGomez2022flamechemistry}, hence increased reactivity due to a higher number of active sites.

Conversely, some impurities may not have a catalytic effect and instead support catalyst deactivation, commonly referred to as poisoning. This has been observed with some halogens, e.g. \ce{Cl} \cite{Hippo1989activesites}, which can be chemisorbed on active sites and prevent further \ce{C}-\ce{O2} reactions \cite{WalkerJr1959carbonoxygenreaction}. The origin of \ce{Cl}-based species can be traced back to the electrolyte used for anodic oxidation, closely related to \ce{Na} as mentioned earlier. In our case, HM63 yielded a higher \ce{Cl} concentration of 18.6$\pm$0.8 ppm \textit{vs.} 4.72$\pm$0.34 and 6.57$\pm$0.42 ppm for AS4 and IM7, respectively. A possible explanation for this difference is that high modulus fibers need higher treatment currents \cite{Morgan2005bookCFs}, which may enhance \ce{Cl} adsorption. The presence of two more halogens was confirmed in the three fibers, with \ce{Br} \& \ce{I} concentrations of 0.0815$\pm$0.0166 \& 0.0440$\pm$0.0056, 0.180$\pm$0.019 \& 0.0781$\pm$0.0072 and 0.064$\pm$0.005 \& 1.54$\pm$0.06 ppm for AS4, IM7 and HM63 fibers, respectively.

Other species were confirmed at very low levels. \ce{Mn} \& \ce{Sb} impurities were detected in the three fibers, i.e., AS4, IM7 and HM63 (0.0447$\pm$0.0029 \& 0.0126$\pm$0.0019, 0.0713$\pm$0.0200 \& 0.0361$\pm$0.0023 and 0.0381$\pm$0.0040 \& 0.0089$\pm$0.00088 ppm, respectively). Other elements where only confirmed in either HM63 (\ce{Ti} at 0.698$\pm$0.107 ppm, \ce{V} at 0.103$\pm$0.004 ppm and \ce{Cu} at 0.150$\pm$0.039 ppm) or AS4 \& IM7 fibers (\ce{Hg} at 0.266$\pm$0.055 and 0.113$\pm$0.057 ppm, respectively). From these elements, \ce{Ti} and \ce{V} have been reported to accelerate \ac{CF} combustion \cite{PatentUS4551487vanadium}, following a similar approach as in other works aiming at preventing fiber fragment release \cite{Sussholz1980fiberrelease, Ganjei1981catalyzedoxidation, Hull1981fibergasification}. From this last group of impurities, the only element that was visually confirmed was \ce{Cu}. Micron-sized \ce{Cu}-based particles were observed via \ac{SEM} and their characteristic effects are discussed in \S \ref{sec:Results_Burnt_Morphology}. We cannot explain the origin of \ce{Cu}, \ce{Hg} and \ce{Sb} impurities detected due to the proprietary nature of \ac{CF} manufacturing process. Nonetheless, further analyses could focus on the synergistic effects that the aforementioned impurities may have \cite{SousaLobo2016kinetics}.

Overall, it is possible that the one-order-of-magnitude difference in \ac{TTF} reported between AS4/IM7 and HM63 in our previous analysis \cite{ChavezGomez2022flamechemistry} may be the result of the latter's lower impurity levels in addition to higher crystallinity and less functional sites (as suggested by the reported modulus and thermal conductivity values \cite{Hexcel_HM63_TDS}). However, since the nature of impurities and their compounds is not precisely known, this hypothesis needs to be tested in future studies.

%%%%%%%%%%%%%%%%%%%%%%%%%%%%%%%%%%%%%%%%%%%%%%%%%
% % % % % % % % % % % % % % % % % % % % % % % % %
%%%%%%%%%%%%%%%%%%%%%%%%%%%%%%%%%%%%%%%%%%%%%%%%%
\subsection{Burnt fibers -- controlled insertion} \label{sec:Results_burnt_fibers}

\subsubsection{Damage morphology} \label{sec:Results_Burnt_Morphology}

\begin{figure}[htbp]
    \centering
    %\captionsetup{justification=centering,margin=2cm}
    \includegraphics[width=\columnwidth,height=0.90\textheight,keepaspectratio]{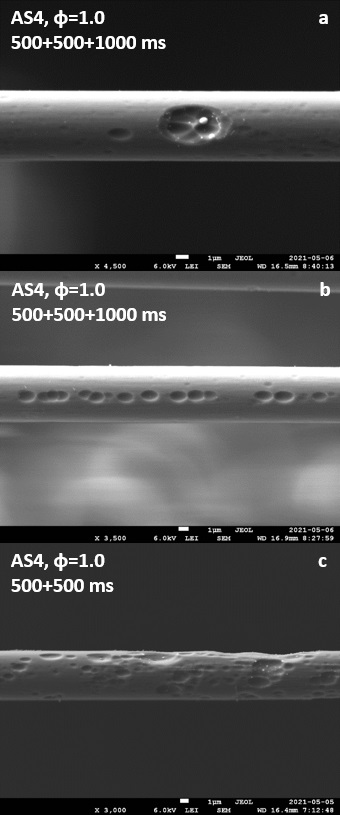}
    \caption{Different types of pits observed on AS4 fibers after exposure to stoechiometric flames ($\phi=1.0$): a) localized damage, b) chain of individual pits and c) random arrangement.}
    \label{fig:Pit_Classification}
\end{figure}

Fig. \ref{fig:Pit_Classification} shows three different \ac{CF} pit group arrangements representative of our \ac{SEM} observations following insertion into flames. These damage morphologies have an oval shape and are typically created by the attack of defects on the basal planes and edge sites of graphite \cite{Rosner1968oxidation, Lee1999defectinduced, Nicholson2005etching}. In our tests, pits were attributed to the presence of buried defects as well as highly reactive species and impurities such as alkali and alkaline earth metals. These impurities can adsorb \ce{O2} dissociatively and readily create oxides, showing strong interaction with the graphene defects and layer edges, usually spreading and having an edge-recession effect \cite{Baker1986catalyst, SousaLobo2016kinetics}. Fig. \ref{fig:Pit_Classification}a shows the results of several pits that coalesced and, due to their proximity, yielded a larger pit. The residues appearing white in \ac{SEM} images correspond to \ce{Ca}-based compounds that initially promoted catalyzed gasification and but then react with \ce{Al}, and \ce{Si} to form stable compounds with weak catalytic activity, as confirmed by \ac{EDS}. In-depth discussion pertaining to the chemical analysis of burnt fibers is presented in \S \ref{sec:Results_EDS_Burnt_fibers}. The second type of pit arrangement is shown in Fig. \ref{fig:Pit_Classification}b as a pit chain which extends on both sides of the field of view. It seems reasonable to attribute this highly-aligned pit formation to the precursor spinning phase. Finally, Fig. \ref{fig:Pit_Classification}c shows a cluster of randomly-placed pits which, as opposed to the aforementioned chain arrangement, could be attributed to post-carbonization surface treatment. In this case, impurities may have remained at random locations on the surface upon electrolyte drying.

\begin{figure}[htbp]
    \centering
    \includegraphics[width=\columnwidth,height=0.90\textheight,keepaspectratio]{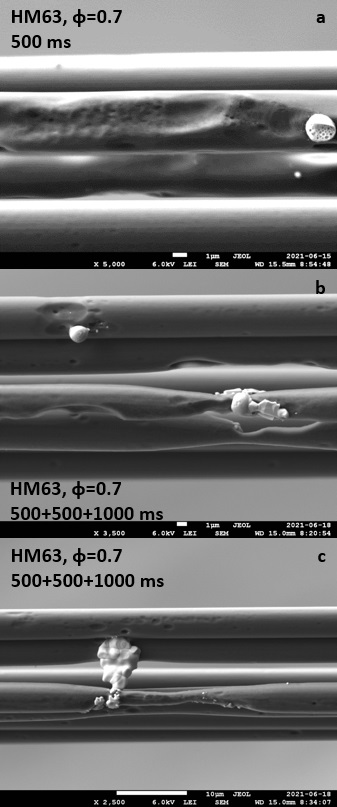}
    \caption{HM63 fibers with mobile \ce{Cu}-based impurities causing channelling (a) and amorphous erosion (b and c).}
    \label{fig:Burnt_HM63_Copper_Damage}
\end{figure}

A second damage mechanism is channelling, as depicted in Fig. \ref{fig:Burnt_HM63_Copper_Damage}a, which shows a channel created by a copper-based macroparticle (confirmed by \ac{EDS}). Channels are the result of mobile metallic impurities, which adsorb oxygen in a non-dissociative manner. These in turn react with the carbon surface without creating stable oxides \cite{Baker1986catalyst, Neeft1998catalytic, SousaLobo2016kinetics} and usually gasify the carbon surface laterally, a pattern commonly described in other carbonaceous materials as worm-like. The trailing path seen in Fig. \ref{fig:Burnt_HM63_Copper_Damage}a shows the high mobility of these impurities, promoting fiber gasification without getting fully oxidized nor deactivated and thus generating extended superficial erosion. In our tests, these mobile impurities likely reach and exceed the \ac{TTa} \cite{Baker1982TammannTemp, SousaLobo2016kinetics}, which is typically described as half of the metal or metal compound's \ac{Tm}. The \ac{TTa} has been interpreted as the temperature where the impurities appear to merge with the carbonaceous surface in a "sintering-like" manner \cite{SousaLobo2016kinetics} without melting, thus ensuring active contact and promoting carbon desorption, bulk diffusion towards the oxidant and final gasification. The contaminant particles detected are therefore likely active catalyzers as the flame temperatures encountered are well above the \ac{TTa} of relevant impurities such as \ce{Cu}-based oxides \ce{CuO} and \ce{Cu2O} (\ac{TTa} $\approx$ \qtylist[list-units = single]{799;752}{\kelvin}, respectively) or \ce{Ca}-based carbonate (\ce{CaCO3}) (\ac{TTa} $\approx$ 549 or \qty{806}{\kelvin}) and highly-reactive oxide (\ce{CaO}) (\ac{TTa} $\approx$ \qty{1422}{\kelvin}) \cite{Mckee1975catalytic, Neeft1998catalytic, Bevilacqua2015CCbrakes}.

The third type of damage observed upon flame attack is amorphous etching. Figs. \ref{fig:Burnt_HM63_Copper_Damage}b and \ref{fig:Burnt_HM63_Copper_Damage}c show HM63 fibers with localized thinning, yielding an hourglass-like shape. This effect is most likely induced by the \ce{Cu}-based impurities next to the damaged areas, as confirmed by \ac{EDS}. Iacocca and Duquette \cite{Iacocca1993catalytic} previously analyzed the catalyzed oxidation of high modulus \ac{PAN}-based \acp{CF} in presence of platinum (\ce{Pt}) in their sample holder. They reported distinct damage mechanisms, namely beaded fiber portions due to uneven thinning as well as fiber splitting, both attributed to changes in fiber morphology. Based on their results and the proximity of \ce{Cu}-based impurities, we attribute this localized thinning to either microstructural changes within the fiber, concentrated impurities, or a combination thereof. Regarding the origin of such \ce{Cu}-based impurities, their size and the associated low \ce{Cu} levels confirmed via \ac{NAA} do not allow the determination of a clear origin. The same rationale applies to \ce{Ti} and \ce{V} impurities confirmed in HM63 fibers (\S \ref{sec:Results_Impurities}), although not discernible upon combined \ac{SEM}/\ac{EDS} analyses. Given that HM63 was the only sized fiber, it is possible that these elements were part of the sizing compound. 

\begin{figure}[htbp]
    \centering
    \includegraphics[width=0.5\textwidth]{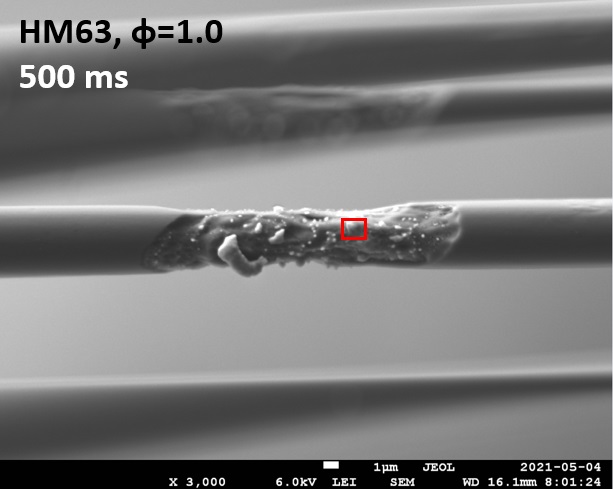}
    \caption{HM63 fiber after \qty{500}{\milli\second} of $\phi=1.0$  flame exposure. See \ac{EDS} spectrum in Fig. \ref{fig:EDS_Burnt_fibers}c.}
    \label{fig:HM63_Amorphous_Damage}
\end{figure}

Fig. \ref{fig:HM63_Amorphous_Damage} shows an HM63 fiber heavily damaged with several residues throughout the affected area, depicting the outcome of catalytic oxidation and poisoning. Amorphous damage spanning $\sim$ \qty{20}{\micro\meter} can be observed where large portions of fiber skin, core or both were etched yielding an heterogeneous central portion, resulting in a similar "hourglass" effect as in Figs. \ref{fig:Burnt_HM63_Copper_Damage}b and \ref{fig:Burnt_HM63_Copper_Damage}c. The red rectangle indicates a region probed with \ac{EDS} and its spectrum is shown in Fig. \ref{fig:EDS_Burnt_fibers}c, confirming the presence of \ce{Ca}, \ce{Mg}, \ce{Si}, \ce{Al} and \ce{P} species. This damage pattern is most likely caused by the concentrated presence of impurities that results in initial violent localized reaction, followed by deactivation and conversion to more stable and less catalytically active species. The residues visible as white specks in Fig. \ref{fig:HM63_Amorphous_Damage} are therefore a mixture of \ce{P}-based compounds, which are known for inhibiting catalytic reactions \cite{Wu2006phosphorusinhibition}, \ce{Ca}- \& \ce{Mg}-based carbonates and/or oxides, well known carbon catalysts \cite{SousaLobo2016kinetics}, along with more stable species such as aluminosilicates. The \ac{EDS} spectra of the largest flake, provided as supplementary material, and of the residues observed on the AS4 fiber, shown in Fig. \ref{fig:EDS_Burnt_fibers}a, show similar elemental signatures and support this hypothesis.

\begin{figure}[htbp]
    \centering
    \includegraphics[width=0.90\columnwidth,height=0.90\textheight,keepaspectratio]{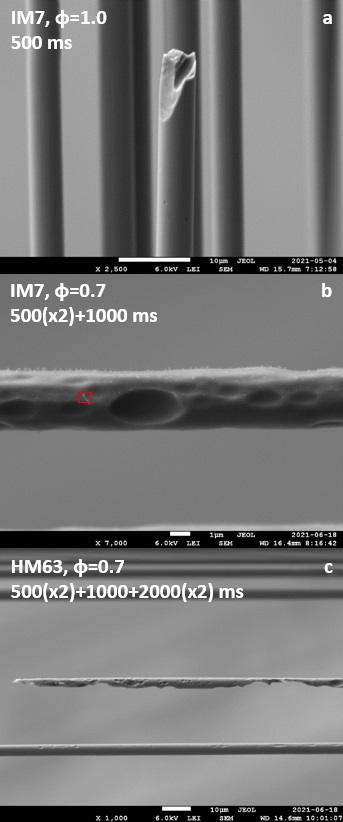}
    \caption{Different damage types: (a) competing pits with no appreciable fiber diameter reduction. (b) Fairly homogeneous pitting and diameter reduction (the \ac{EDS} spectrum of the red rectangle is shown in Fig. \ref{fig:EDS_Burnt_fibers}b). (c) Heavy heterogeneous damage after long exposure.}
    \label{fig:fiber_Failure}
\end{figure}

If no external force is exerted, fiber rupture can be caused by small mechanical loads induced by the flame itself, fiber diameter reduction or pore/pit growth and subsequent coalescence. Fig. \ref{fig:fiber_Failure}a shows an IM7 fiber which failed after only \qty{500}{\milli\second} of stoichiometric flame exposure due to collaborative pit growth. Despite the quiescent nature of the flame, the flow exerted a small yet sufficient force to bend the fiber, causing fracture. Much slower and more homogeneous diameter reduction was also observed for the same fiber, as shown in Fig. \ref{fig:fiber_Failure}b, despite being exposed for a longer period (\qty{2000}{\milli\second}) to a fuel-lean flame ($\phi=0.7$) with a more aggressive oxidizing atmosphere. The \ac{EDS} spectrum of this slowly degrading fiber location reveals less abundant impurities, without the presence of \ce{Ca},  as shown in Fig. \ref{fig:EDS_Burnt_fibers}b, clearly demonstrating the dominating role of a limited number of catalytically active species in the fiber failure mechanisms. Fig. \ref{fig:fiber_Failure}c shows an HM63 fiber with heterogeneous damage caused by pitting, channelling and amorphous erosion after \qty{8000}{\milli\second} of accumulated exposure.

%%%%%%%%%%%%%%%%%%%%%%%%%%%%%%%%%%%%%%%%%%%%%%%%%
% % % % % % % % % % % % % % % % % % % % % % % % %
%%%%%%%%%%%%%%%%%%%%%%%%%%%%%%%%%%%%%%%%%%%%%%%%%
\subsubsection{Surface \& pit chemical analysis} \label{sec:Results_EDS_Burnt_fibers}

\begin{figure}[htbp]
    \centering
    \includegraphics[width=\textwidth,height=0.90\textheight,keepaspectratio]{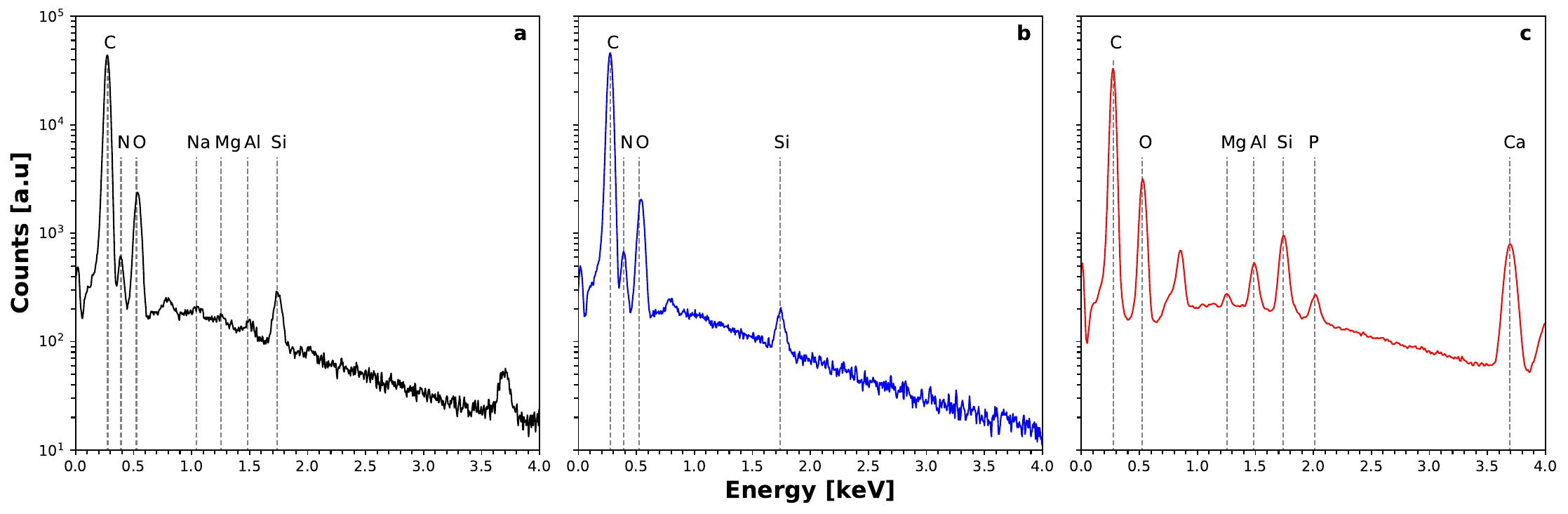}
    \caption{Examples of \ac{EDS} spectra from burnt fibers: a) AS4 (Fig. \ref{fig:AS4_IM7_Pit_Sequence}c), b) IM7 (Fig. \ref{fig:fiber_Failure}b) and c) HM63 (Fig. \ref{fig:HM63_Amorphous_Damage}) fiber surfaces. All confirmed peaks correspond to the K$_{\alpha}$ values.  The plots are shown in an unconventional semi-log scale for clarity of peaks above $\sim$\qty{1.0}{\kilo\eV}.}
    \label{fig:EDS_Burnt_fibers}
\end{figure}

Fig. \ref{fig:EDS_Burnt_fibers} shows the \ac{EDS} spectra representative of the three fibers types after flame exposure. Again, the y-axis has been transformed to allow the identification of weak signals caused by minuscule impurities. Otherwise, in a linear scale, they are typically indiscernible in comparison with the dominating signatures of \ce{C}, \ce{N} and \ce{O}. Fig.\ref{fig:AS4_IM7_Pit_Sequence}c shows a small red rectangle enclosing a round white particle, with the corresponding \ac{EDS} spectrum shown in Fig. \ref{fig:EDS_Burnt_fibers}a. It can be observed that HM63 fibers do not show a \ce{N} peak. This confirms that the higher \ac{HTT} needed for high modulus fibers also help to completely remove the \ce{N} atoms from the original \ac{PAN} structure. The residues observed in large pits yielded well-defined \ce{Si} and \ce{Ca} peaks, along with small contributions from \ce{Na}, \ce{Mg} and \ce{Al} atoms. IM7 fibers are an exception here, as no impurities residues appearing as white particles in SEM images could be observed and EDS spectra lacked \ce{Ca}, \ce{Al} and \ce{Mg} peaks. The relative particle immobility suggests that, upon pit creation, the aforementioned alkali and alkaline earth metals reacted with \ce{Si} and \ce{Al} species forming stable aluminosilicates from otherwise reactive elements \cite{Arnold2019catalysts}. Although it has been suggested that "inert" oxides (e.g. \ce{TiO2} or \ce{Al2O3}) may mechanically erode graphite layers in other oxidative conditions \cite{Chang1991scanning}, here we attribute the genesis of pits to \ce{Ca}- and, to some extent, \ce{Na}- and \ce{Mg}-based species. Upon reaction with \ce{Al} and \ce{Si} atoms, more stable species were likely created thus preventing further catalytic gasification. 
 
%%%%%%%%%%%%%%%%%%%%%%%%%%%%%%%%%%%%%%%%%%%%%%%%%
% % % % % % % % % % % % % % % % % % % % % % % % %
%%%%%%%%%%%%%%%%%%%%%%%%%%%%%%%%%%%%%%%%%%%%%%%%%
\subsubsection{Damage evolution} \label{sec:Results_Burnt_Damage_Evolution}

\begin{figure}[htbp]
    \centering
    \includegraphics[width=0.99\columnwidth,height=0.90\textheight,keepaspectratio]{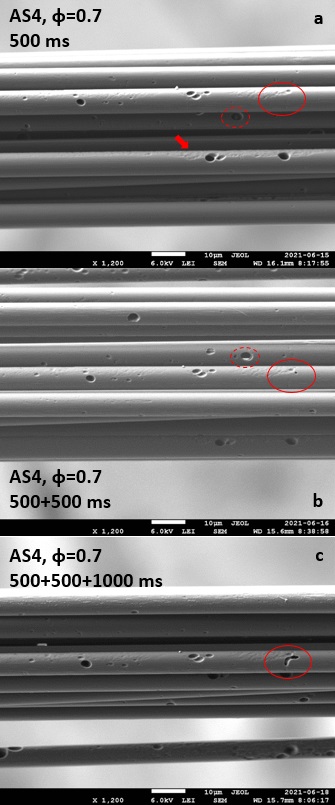}
    \caption{Sequential exposure of AS4 fibers ($\phi$=0.7), showing the evolution of a channel caused by a mobile impurity (solid-line oval): absent pit (a), pit genesis (b) and transition into a fully developed channel (c).}
    \label{fig:Burnt_AS4_Channelling}
\end{figure}

The sequential flame insertion approach implemented, followed by repeated fixed-point SEM analysis, allowed the observation of damage evolution resulting from controlled flame exposure.  Fig. \ref{fig:Burnt_AS4_Channelling} shows the genesis of a pit and subsequent channelling effect of mobile impurities, indicated by the solid line oval. The right side of Fig. \ref{fig:Burnt_AS4_Channelling}a shows a small region with amorphous damage, with a large pit on the right, created after a \qty{500}{\milli\second} flame exposure. After an additionnal \qty{500}{\milli\second} insertion, a new small pit appears on the lower right side of the existing pit, as shown in Fig. \ref{fig:Burnt_AS4_Channelling}b. This is the starting point of the channel, and can be attributed to an impurity that remained buried and did not react in the first flame exposure. Finally, Fig. \ref{fig:Burnt_AS4_Channelling}c shows the resulting angled channel after a total of \qty{2000}{\milli\second} in the flame atmosphere. 

Most other features on the same fiber did not grow as fast as the channel, confirming that most catalyst particles were quickly deactivated or removed following initial flame exposure. Amorphous erosion in form of skin peeling was observed in a few locations, as indicated by an arrow in \ref{fig:Burnt_AS4_Channelling}a, but did not appear to evolved following subsequent flame insertions. Defective carbon structure combined with interleaved impurities could have promoted this exfoliation \cite{Yoshida1990exfoliation, Serp1997vaporgrownCF}. However, a tunnel created by a mobile impurity going completely through a fiber is indicated in \ref{fig:Burnt_AS4_Channelling}a and \ref{fig:Burnt_AS4_Channelling}b by the dashed-line oval. 

\begin{figure}[htbp]
    \centering
    \includegraphics[width=0.5\columnwidth,height=0.90\textheight,keepaspectratio]{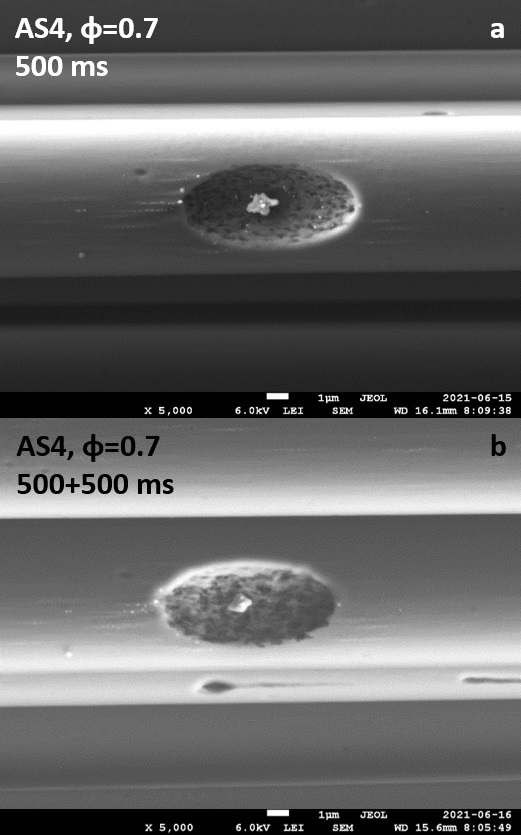}
    \caption{Pit and internal porosity evolution of an AS4 fiber upon sequential flame exposure ($\phi$=0.7).}
    \label{fig:AS4_Internal_Pore_Porosity_Evolution}
\end{figure}

\begin{figure}[htbp]
    \centering
    \includegraphics[width=0.99\textwidth,height=0.90\textheight,keepaspectratio]{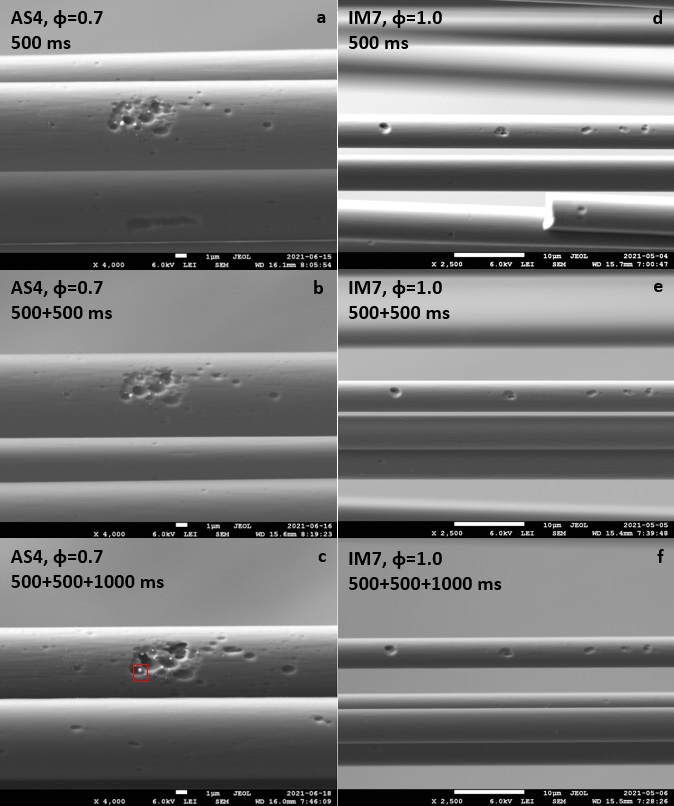}
    \caption{Pit evolution of AS4 (a-c) and IM7 (d-f) fibers after sequential exposure to lean ($\phi=0.7$) and stoichiometric ($\phi=1.0$) \ce{CH4}/air flames, respectively. Rectangle at (c) shows the probed area of the \ac{EDS} spectrum shown in Fig. \ref{fig:EDS_Burnt_fibers}a.}
    \label{fig:AS4_IM7_Pit_Sequence}
\end{figure}

Fig. \ref{fig:AS4_Internal_Pore_Porosity_Evolution} shows the rapid porosity evolution within an extremely large pit created on the surface of an AS4 fiber during the first half second of exposure to the flame. The pit is most likely due to a mixture of \ce{Ca}-, \ce{Na}- and/or \ce{Mg}-based species which reacted with \ce{Al} and/or \ce{Si}, as suggested by \ac{EDS} analyses of similar residues (e.g. Figs. \ref{fig:HM63_Amorphous_Damage} and \ref{fig:AS4_IM7_Pit_Sequence}c), as discussed in \S\S \ref{sec:Results_Burnt_Morphology} and \ref{sec:Results_EDS_Burnt_fibers}. The pit's long axis dimension exceeds the fiber diameter itself, with an apparent pit growth rate ($\sim$\qty{14800}{\nano\meter\per\second}) that largely exceeds the average pit growth rates observed under the same conditions. After the second insertion, the pit did not show a noticeable change of dimensions. Moreover, its periphery shows heterogeneous damage with several dents and half pits with apparent growth rates in the \qtyrange[range-phrase=--,range-units=single]{400}{1000}{\nano\meter\per\second} range. On the other hand, the smaller pit located on the upper left side yields a moderate and more constant growth rate of $\sim$\qty{115}{\nano\meter\per\second}. This reveals a very large variability in growth rates for extremely large pits, whereas the growth rate of submicron surface pits appears more homogeneous. Although it was possible to identify certain elements at selected damaged locations via \ac{EDS}, their effect on pit growth rates cannot be determined since different compounds from the same metallic impurity can promote catalysis to different extents, as well synergistic or poisoning effects in the presence of two or more catalysts \cite{SousaLobo2016kinetics}. Knowledge of the original location of impurities as well as of the nature of their compounds is therefore needed.

To evaluate the effect of flame chemistry on pit evolution, the three fibers were sequentially inserted into the three types of flame. Unfortunately, it was only possible to follow up the same pits or channels on two out of the three fibers, i.e., AS4 and IM7. The pit growth rates reported here for these two fiber types were also obtained in different flame conditions, fuel-lean ($\phi=0.7$) for AS4 and stoichiometric ($\phi=1.0$) for IM7. Figs. \ref{fig:AS4_IM7_Pit_Sequence}a-c show the evolution of a cluster that appeared upon the first \qty{500}{\milli\second} insertion on the surface of an AS4 fiber, with coalescence of individual pits upon subsequent insertions. Fig. \ref{fig:AS4_IM7_Pit_Sequence}d shows IM7 fibers with pits created during the first exposure to stoichiometric flames. These pits did not change significantly in size following sequential insertion as shown in Figs. \ref{fig:AS4_IM7_Pit_Sequence}e-f. 

\begin{figure}[htbp]
    \centering
    \includegraphics[width=\textwidth,height=0.90\textheight,keepaspectratio]{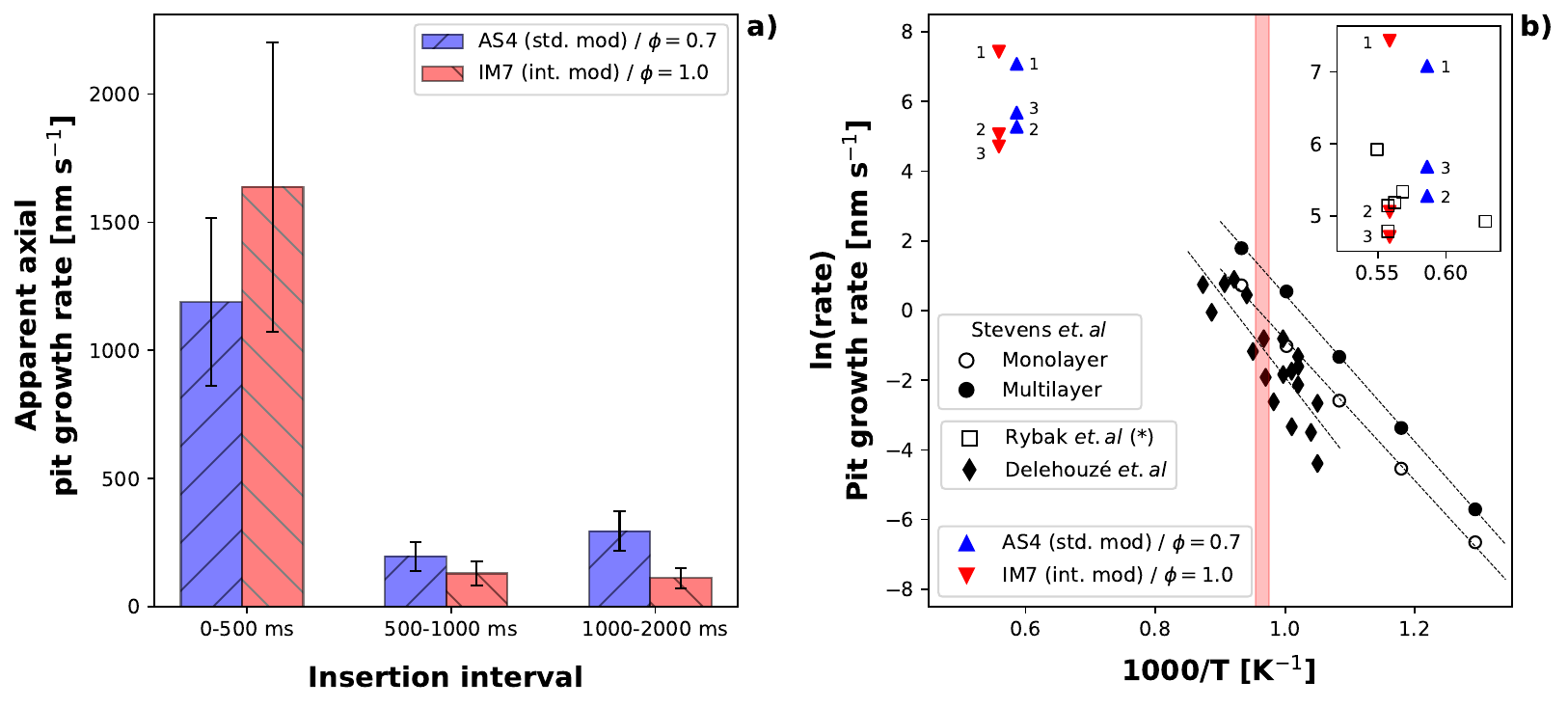}
    \caption{(a) Apparent axial pit growth rate \textit{vs.} insertion interval. The means (AS4, $n=18$; IM7 $n=13$) were obtained from randomly-chosen pits shown in Fig. \ref{fig:AS4_IM7_Pit_Sequence}. Error bars indicate the 95\% \ac{CI}. (b) Arrhenius plot comparing the pit growth rates from a), i.e. AS4 (\textcolor{blue}{\FilledSmallTriangleUp}) and IM7 (\textcolor{red}{\FilledSmallTriangleDown}) \textit{vs.} data from Stevens \textit{et al.} ($\bullet$,$\circ$) \cite{Stevens1998kinetics} \textcolor{blue}{(unknown \ce{O2} partial pressure; unrestricted air flow at room pressure using a tube furnace)} and Delehouzé \textit{et al.} ({\scriptsize $\blacklozenge$}) \cite{Delehouze2011transition} \textcolor{blue}{(pure \ce{O2} at \qty{140}{\pascal} using a \ac{HTSEM})}. The vertical strip indicates the hexagonal-circular pit transition zone from \cite{Delehouze2011transition}. Markers 1, 2, 3 correspond to the \qtyrange[range-phrase=--,range-units=single]{0}{500}{\milli\second}, \qtyrange[range-phrase=--,range-units=single]{500}{1000}{\milli\second}, and \qtyrange[range-phrase=--,range-units=single]{1000}{2000}{\milli\second} insertion intervals, respectively. The inset compares the values from (a) \textit{vs.} carbon black diameter reduction rate in ethylene flames from Rybak \textit{et al.} ({\scriptsize $\square$}) \cite{Rybak1995oxidation} \textcolor{blue}{(ethylene/air flame at \qty{100}{\kilo\pascal} using a custom burner setup)}.}
    \label{fig:SEM_Pit_Growth}
\end{figure}

To gain insight into the role played by flame chemistry and fiber microstructure, the apparent pit growth rates were calculated for all exposure intervals. The results are reported in Fig. \ref{fig:SEM_Pit_Growth}a for random pits found on AS4 ($n=18$, from Figs.\ref{fig:AS4_IM7_Pit_Sequence}a-c) and IM7 fibers ($n=13$, from Figs.\ref{fig:AS4_IM7_Pit_Sequence}d-f). Pits were only measured lengthwise to avoid parallax errors induced in widths due to fiber curvature. Most of these pits tend to grow faster lengthwise, following crystallite orientation, which in turn is a function of \acp{CF} \ac{HTT} \cite{Takaku1990structure}. When fibers did not break nor were hidden by other fibers due to flame-induced rearrangement, the same pits were measured after successive insertions. For each step, a mean apparent growth rate and its 95\% \ac{CI} were calculated considering all pits. During the first \qty{500}{\milli\second} insertion, rapid pit nucleation took place in both fuel-lean and stoichiometric flames, yielding 1189$\pm$326 and 1637$\pm$\qty{563}{\nano\meter\per\second} for AS4 and IM7 fibers, respectively. Upon a second \qty{500}{ms} insertion, the growth rates decreased by an order-of-magnitude, indicating a major reduction of impurity-induced catalytic effects in these specific areas. It is possible that some impurities remained active, although to a lesser extent. Henceforth, in absence of catalysts, pit growth would have been driven by the oxidative species present in the flame atmosphere, as well as by the amount and nature of active sites presented by the fiber microstructure. Both flames yielded an apparent pit growth rate of 196 $\pm$ 57 and 129 $\pm$ \qty{48}{\nano\meter\per\second} for AS4 and IM7 fibers, respectively. The third and final insertion lasted for \qty{1000}{\milli\second}, clearly revealing the immediate effect of flame chemistry. The axial pit growth rate of 294 $\pm$ \qty{78}{\nano\meter\per\second} measured for AS4 fibers is much larger than the 111 $\pm$ \qty{40}{\nano\meter\per\second} of IM7 fibers. This difference is in contrast to the \ac{TGA}/\ac{DSC} data discussed in our previous work \cite{ChavezGomez2022flamechemistry}, which indicates a higher reactivity of IM7 fibers \textit{vs.} AS4 in air. After the third insertion, the pit growth in IM7 fibers remains esentially unchanged, suggesting a stable attack by the stoichiometric flame. However, the fiber (AS4) exposed to the lean flame atmosphere, where significantly more \ce{O2} is present, shows an increase in growth rate. 

Unfortunately, it was not possible to follow-up these trends over long durations due to fiber breakage. Moreover, although the high modulus fibers (HM63) showed similar damage mechanisms to those observed on AS4 and IM7 fibers, pit growth rates are not shown in Fig. \ref{fig:SEM_Pit_Growth}a since it was not possible to follow up any oxidation-induced features throughout after each sequence.

Considering our test conditions and the definition of macroporosity, i.e., $\gtrsim$ \qty{50}{\nano\meter} \cite{Rouquerol1994IUPACpores}, it seems reasonable to assume enhanced \ac{CF} reactivity in presence of macropores in such reactive environment. In the case of \acp{CF} with well-defined core-sheath structure, i.e. standard (AS4) and intermediate (IM7) modulus fibers, large pits will reveal a larger portion of the amorphous core, increasing the fiber reactivity owing to the inner imperfect structure. In other words, surface pitting contributes toward more-developed internal pore networks, which in turn can translate into accelerated fiber failure even under ideal low external forces due to percolative fragmentation \cite{Fuertes1996gasification, Salatino1993percolative}. Other factors may come at play affecting fiber failure, for instance the flow regime, e.g. Knudsen and transition flow, as well as the Thiele modulus, as pointed out by other researchers \cite{Panerai2013flow, Panerai2019experimentaloxidation}. More research is needed to evaluate the evolution of the internal pore network and \ac{ASA} and their effect on the gasification rate, i.e. oxidation kinetics. These analyses could be aided by adsorption or small-angle scattering techniques.

\subsubsection{Pit shape \& growth rate comparison} 

An Arrhenius plot is presented to put the values discussed in \S\ref{sec:Results_Burnt_Damage_Evolution} into perspective. Fig. \ref{fig:SEM_Pit_Growth}b contains the pit growth rates of AS4 (\textcolor{blue}{\FilledSmallTriangleUp}) and IM7 (\textcolor{red}{\FilledSmallTriangleDown}) fibers from Fig. \ref{fig:SEM_Pit_Growth}a marked as 1, 2 and 3, which correspond to the apparent rates obtained from the \qtyrange[range-phrase=--,range-units=single]{0}{500}{}, \qtyrange[range-phrase=--,range-units=single]{500}{1000}{}, and \qtyrange[range-phrase=--,range-units=single]{1000}{2000}{\milli\second} insertion intervals, respectively. For comparison purposes, other pit diameter growth rates are shown, i.e., values from Stevens \textit{et al.} ($\bullet$,$\circ$) \cite{Stevens1998kinetics} and Delehouzé \textit{et al.} ({\scriptsize $\blacklozenge$}) \cite{Delehouze2011transition}. Both reference data sets were obtained from \ac{HOPG} surfaces at similar length scales (from hundreds of \qty{}{\nano\meter} up to a few microns), after \ce{O2} attack under different experimental conditions: unrestricted air flow at room pressure using a tube furnace with unknown \ce{O2} partial pressure \cite{Stevens1998kinetics}, and pure \ce{O2} at \qty{140}{\pascal} using a \ac{HTSEM} \cite{Delehouze2011transition}. A vertical red strip indicates the hexagonal-circular pit transition zone ($\sim$\qtyrange[range-phrase=--,range-units=single]{1025}{1050}{\kelvin}) also reported by Delehouzé \textit{et al.} \cite{Delehouze2011transition}. Under pure \ce{O2} and low pressure (\qty{140}{\pascal}) conditions, they found that hexagonal pits were formed below this temperature threshold, while round features were favoured at higher temperatures. Two characteristic trends can be identified upon examination of Fig. \ref{fig:SEM_Pit_Growth}b and comparison between different datasets, namely temperature-driven pit geometry and apparent growth rate.

Regarding pit geometry, our oxidation tests were carried at a position that ensures flame temperatures at least $\sim$\qty{650}{\kelvin} above the transition threshold \cite{Delehouze2011transition}, hence hexagonal pitting was not expected. It is extremely unlikely that hexagonal pits form on the \ac{CF} surface. Only round pitting has been reported in other \ac{CF}-based works, regardless of the etching conditions \cite{Sussholz1980fiberrelease, Ismail1991reactivity, Halbig2008oxidation, Tong2011kineticstensile, Bertran2013oxidation, Kachold2016preheating, Marshall1991topography, Hoffman1992scanning, PittmanJr1997nitricacid, Serp1997vaporgrownCF, Cho1997protection, Tzeng2010resistance, Barnet1973oxygenplasma, Iacocca1993catalytic, Panerai2013flow, Panerai2019experimentaloxidation, Cochell2021nanoscaleoxidation, ChavezGomez2019fiberoxidation, Cho1993microscopicablation, Eibl2017respirablefibers, Bell1979fiberburning}, either as well-defined individual features, pit clustering and subsequent coalescence, or seemingly amorphous. This is explained by the heterogeneity of \ac{PAN}-based \acp{CF} structure, highly disordered compared to graphite models, e.g. \ac{HOPG} or natural graphite. Moreover, it has been suggested that oxidative etching of \ac{CF} surfaces results in the removal of full crystallites \cite{Marshall1991topography} which have a size on the order of a few nanometers for the \acp{CF} studies here \cite{ChavezGomez2022flamechemistry}. Highly oxidizing environments also do not yield hexagonal pitting since highly-energetic species such as radicals or atomic species are able to attack the graphitic structure regardless of the type of site, i.e. edge site or on the basal plane. For instance, Nicholson \textit{et al.} \cite{Nicholson2005etching} showed that round pitting was formed on \ac{HOPG} surfaces under attack of highly reactive species (\ce{O}($^3$P)/\ce{O2} mixture) even at rather mild temperatures ($\sim$\qtyrange[range-phrase=--,range-units=single]{298}{423}{\kelvin}). 

Values obtained for AS4 and IM7 fibers range from hundreds to thousands of \qty{}{\nano\meter\per\second}, whereas the reference etch rates from \cite{Stevens1998kinetics, Delehouze2011transition} remain below $\sim$\qty{10}{\nano\meter\per\second}. As pointed out by Blyholder \textit{et al.} \cite{Blyholder1958kinetic} and Stevens \textit{et al.} \cite{Stevens1998kinetics}, direct comparison between homogeneous surface oxidation and pit growth rates is a challenge owing to different test conditions, even when similar materials are considered owing to the different processing conditions of different carbonaceous materials. Since none of the reference works considered fire conditions, a flame-based benchmark was defined. Diameter change rates of carbon black oxidized in ethylene flame reported by Rybak \textit{et al.} ({\scriptsize $\square$}) \cite{Rybak1995oxidation} are shown in the inset for direct comparison with our pit growth rates. Both datasets are in the same range ($\sim$\qtyrange[range-phrase=--,range-units=single]{100}{400}{\nano\meter\per\second}) except for the values obtained after initial exposure due to catalytic effects (marked as 1). Despite the challenge of comparing the reactivity of dissimilar carbonaceous materials under different oxidizing conditions, the similarity between our data and carbon black oxidation in flames is encouraging. Pit growth rates have not previously been reported for \acp{CF} surfaces under highly-reactive conditions, let alone from direct exposure to flame conditions. We consider that the values reported here are a good starting point for the validation of damage models intended for \ac{CF}-based materials in combustion environments representative of fire hazards.

%%%%%%%%%%%%%%%%%%%%%%%%%%%%%%%%%%%%%%%%%%%%%%%%%
% % % % % % % % % % % % % % % % % % % % % % % % %
%%%%%%%%%%%%%%%%%%%%%%%%%%%%%%%%%%%%%%%%%%%%%%%%%
\section{Conclusions}

The oxidative behavior of \acp{CF} under open flame attack was studied using an original technique consisting in time-controlled insertion of three different types of unloaded \acp{CF} into flames and \textit{post hoc} fixed point observations. This enabled sequential damage assessment and precise follow-up on fire-induced damages. Localized pitting and homogeneous fiber diameter reduction were both observed, with our investigation focusing on the former. Analytical methods were implemented to precisely identify and quantify impurities (\ac{NAA}, \ac{SEM}, and \ac{EDS}), before and after flame exposure. Pitting was driven by the presence of metallic impurities, namely alkali and alkaline earth metals, that catalyzed the oxidation process. Some impurities remain active for several seconds and had a channelling effect owing to their mobility (e.g. \ce{Cu}-based). Other were found to be quickly deactivated or removed following the first flame exposure, such as \ce{Na}, \ce{Mg} and \ce{Ca}. Intense pit nucleation and growth were observed for all fiber types during the first \qty{500}{\milli\second} of flame exposure, with apparent diameter growth rates of 1189$\pm$326 and 1637$\pm$\qty{563}{\nano\meter\per\second} for AS4 and IM7 fibers, respectively. After this initial period, the \ac{CF} degradation rate drops by an order of magnitude and appears to be controlled by flame chemistry rather than the presence of impurities, with fuel-lean flames resulting in slightly faster pit growth rates from their increased \ce{O2} content. The pit growth rates were compared with the literature which considers different experimental conditions. Extrapolation from such works is not possible given the difference in materials, temperatures and the concentration of oxygen or other reactive species, e.g., radicals. However, our results are coherent with these literature data, since they are in the same order of magnitude reported in other flame-based works.

Although impurities were assessed quantitatively by \ac{NAA}, it was not possible to determine the effect of specific elements on pit growth rates. Impurities were qualitatively identified via \ac{EDS} at certain damaged locations, but the nature of their original compounds remains unknown, i.e., whether metals were found as carbonates, acetates, oxides, etc. Precise knowledge of the spatial location of impurities and of the nature of their compounds is needed in future studies aiming to model their effect on the thermomechanical behavior of \acp{CF}.

Our results highlight that flame chemistry, microstructure, and impurities are all key parameters in \ac{CF} oxidation, with the former often overlooked in the assessment of \ac{CF} fire properties. On a more practical side, we expect that this works helps to improve the fire safety of \ac{CF}-based structures, since current models often assume homogeneous diameter reduction. Moreover, fire resistance regulations tend to disregard the importance of the flame stoichiometry, considering it an outcome of flame calibration focusing on temperature and heat flux. Using a simple gaseous fuel (\ce{CH4}), we have demonstrated that the composition of the reactive atmosphere needs to be given careful consideration in standardized fire tests, but also in other highly reactive conditions where carbon gasification is of interest.

%%%%%%%%%%%%%%%%%%%%%%%%%%%%%%%%%%%%%%%%%%%%%%%%%
% % % % % % % % % % % % % % % % % % % % % % % % %
%%%%%%%%%%%%%%%%%%%%%%%%%%%%%%%%%%%%%%%%%%%%%%%%%

\section{Acknowledgments}
The authors want to acknowledge the financial support from the Natural Sciences and Engineering Research Council of Canada (NSERC/CRSNG) grant no. RGPIN-06410-2016 (NSERC Discovery). Pablo Chávez-Gómez is extremely grateful to the National Science and Technology Council of Mexico (CONACYT) for his doctoral scholarship. The authors also thank Dr. Martin Luckabauer for his valuable feedback, as well as Max Thouin and Eddie Rubey from Hexcel Corp. for supplying some of the samples used in the study.

\end{spacing}

\bibliographystyle{ieeetr}
\bibliography{References.bib}

\begin{thebibliography}{100}

\bibitem{DOT-FAA-AR-TN11-8improvements}
C.~Sarkos, ``{Improvements in Aircraft Fire Safety Derived From FAA Research
  Over the Last Decade},'' tech. rep., {Federal Aviation Administration}, May
  2011.
\newblock {DOT/FAA/AR-TN11/8}.

\bibitem{Mouritz2006book}
A.~P. Mouritz and A.~G. Gibson, {\em {Fire Properties of Polymer Composite
  Materials}}, vol.~143.
\newblock Springer Science \& Business Media, 2006.

\bibitem{Langot2021multiphysics}
J.~Langot, T.~Pelzmann, P.~Chávez-Gómez, C.~S. Boanta, J.~Karsten,
  B.~Fiedler, M.~Lévesque, and E.~Robert, ``{Multi-physics modeling of the
  ignition of polymer matrix composites (PMCs) exposed to fire},'' {\em Fire
  Safety Journal}, vol.~122, p.~103312, June 2021.

\bibitem{Langot2021reactionscheme}
J.~Langot, P.~Chávez-Gómez, M.~Lévesque, and E.~Robert, ``{Modeling the
  thermal decomposition and residual mass of a carbon fiber epoxy matrix
  composite with a phenomenological approach: effect of the reaction scheme},''
  {\em Fire and Materials}, vol.~46, pp.~262--276, 2022.

\bibitem{Hull1981fibergasification}
G.~Hull, C.~R. Glazer, J.~Harper-Tervet, M.~F. Humphrey, and K.~N. Ramohalli,
  ``{Gasification of Carbon Fiber Composites for the Alleviation of Electrical
  Hazards},'' {\em Industrial \& Engineering Chemistry Product Research and
  Development}, vol.~20, no.~3, pp.~552--555, 1981.

\bibitem{Ganjei1981catalyzedoxidation}
J.~Ganjei, D.~Dominguez, J.~Mackey, and J.~Murday, ``Catalyzed combustion of
  carbon fibers from carbon fiber-resin composites,'' tech. rep., {Naval
  Research Laboratory, Washington D.C.}, 1981.
\newblock {NRL Memorandum Report 4486}.

\bibitem{Sussholz1980fiberrelease}
B.~Sussholz, ``{Evaluation of Micron Size Carbon Fibers Released from Burning
  Graphite Composites},'' tech. rep., National Aeronautics and Space
  Administration, April 1980.
\newblock {NASA-CR-159217}.

\bibitem{Gandhi1999potential}
S.~Gandhi, R.~Lyon, and L.~Speitel, ``{Potential health hazards from burning
  aircraft composites},'' {\em Journal of Fire Sciences}, vol.~17, no.~1,
  pp.~20--41, 1999.

\bibitem{Mouritz2009toxicity}
A.~P. Mouritz, ``{Review of Smoke Toxicity of Fiber-Polymer Composites Used in
  Aircraft},'' {\em Journal of Aircraft}, vol.~46, no.~3, pp.~737--745, 2009.

\bibitem{WalkerJr1959carbonoxygenreaction}
F.~Walker~Jr, P. L.and Rusinko~Jr. and L.~Austin, ``{Gas Reactions of
  Carbon},'' in {\em Advances in Catalysis} (D.~Eley, P.~Selwood, and P.~B.
  Weisz, eds.), vol.~XI, pp.~133--221, Academic Press, 1959.

\bibitem{Ismail1991reactivity}
I.~M.~K. Ismail, ``{On the reactivity, structure, and porosity of carbon fibers
  and fabrics},'' {\em Carbon}, vol.~29, no.~6, pp.~777--792, 1991.

\bibitem{Hippo1989activesites}
E.~J. Hippo, N.~Murdie, and A.~Hyjazie, ``{The role of active sites in the
  inhibition of gas-carbon reactions},'' {\em Carbon}, vol.~27, no.~5,
  pp.~689--695, 1989.

\bibitem{Figueiredo2010surfacechemistry}
J.~L. Figueiredo and M.~F.~R. Pereira, ``{The role of surface chemistry in
  catalysis with carbons},'' {\em Catalysis Today}, vol.~150, no.~1-2,
  pp.~2--7, 2010.

\bibitem{Terrones2010graphene}
M.~Terrones, A.~R. Botello-M\'{e}ndez, J.~Campos-Delgado,
  F.~L\'{o}pez-Ur\'{i}as, Y.~I. Vega-Cant\'{u}, F.~J. Rodr\'{i}guez-Mac\'{i}as,
  A.~L. El\'{i}as, E.~Mu\~{n}oz Sandoval, A.~G. Cano-M\'{a}rquez, J.-C.
  Charlier, and H.~Terrones, ``{Graphene and graphite nanoribbons: Morphology,
  properties, synthesis, defects and applications},'' {\em Nano Today}, vol.~5,
  no.~4, pp.~351--372, 2010.

\bibitem{Ismail1987activesurface}
I.~M.~K. Ismail, ``{Structure and active surface area of carbon fibers},'' {\em
  Carbon}, vol.~25, no.~5, pp.~653--662, 1987.

\bibitem{Lee1999defectinduced}
S.~M. Lee, Y.~H. Lee, Y.~G. Hwang, J.~R. Hahn, and H.~Kang, ``{Defect-Induced
  Oxidation of Graphite},'' {\em Physical Review Letters}, vol.~82, no.~1,
  pp.~217--220, 1999.

\bibitem{Hahn1999mechanistic}
J.~R. Hahn, H.~Kang, S.~M. Lee, and Y.~H. Lee, ``{Mechanistic Study of
  Defect-Induced Oxidation of Graphite},'' {\em The Journal of Physical
  Chemistry B}, vol.~103, no.~45, pp.~9944--9951, 1999.

\bibitem{Hahn2005kinetic}
J.~R. Hahn, ``{Kinetic study of graphite oxidation along two lattice
  directions},'' {\em Carbon}, vol.~43, no.~7, pp.~1506--1511, 2005.

\bibitem{Morgan2005bookCFs}
P.~Morgan, {\em {Carbon Fibers and their Composites}}.
\newblock CRC Press, 2005.

\bibitem{Newcomb2016carbonfibers}
B.~A. Newcomb, ``{Processing, structure, and properties of carbon fibers},''
  {\em {Composites Part A: Applied Science and Manufacturing}}, vol.~91, no.~3,
  pp.~262--282, 2016.

\bibitem{Perret1970microstructure}
R.~Perret and W.~Ruland, ``{The Microstructure of PAN-Base Carbon Fibres},''
  {\em Journal of Applied Crystallography}, vol.~3, no.~6, pp.~525--532, 1970.

\bibitem{Barnet1973oxygenplasma}
F.~R. Barnet and M.~K. Norr, ``{Carbon fiber etching in an oxygen plasma},''
  {\em Carbon}, vol.~11, no.~4, pp.~281--288, 1973.

\bibitem{Guigon1984highstrength}
M.~Guigon, A.~Oberlin, and G.~Desarmot, ``{Microtexture and Structure of Some
  High Tensile Strength, PAN-Base Carbon Fibres},'' {\em Fibre Science and
  Technology}, vol.~20, no.~1, pp.~55--72, 1984.

\bibitem{Guigon1984highmodulus}
M.~Guigon, A.~Oberlin, and G.~Desarmot, ``{Microtexture and Structure of Some
  High-modulus, PAN-Base Carbon Fibres},'' {\em Fibre Science and Technology},
  vol.~20, no.~3, pp.~177--198, 1984.

\bibitem{Shi2021oxidation}
L.~Shi, M.~Sessim, M.~R. Tonks, and S.~R. Phillpot, ``{High-temperature
  oxidation of carbon fiber and char by molecular dynamics simulation},'' {\em
  Carbon}, vol.~185, pp.~449--463, 2021.

\bibitem{Halbig2008oxidation}
M.~C. Halbig, J.~D. McGuffin-Cawley, A.~J. Eckel, and D.~N. Brewer,
  ``{Oxidation Kinetics and Stress Effects for the Oxidation of Continuous
  Carbon Fibers within a Microcracked C/SiC Ceramic Matrix Composite},'' {\em
  Journal of the American Ceramic Society}, vol.~91, no.~2, pp.~519--526, 2008.

\bibitem{Kim2011commercial}
D.-H. Kim, B.-H. Kim, K.-S. Yang, Y.-H. Bang, S.-R. Kim, and H.-K. Im,
  ``{Analysis of the Microstructure and Oxidation Behavior of Some Commercial
  Carbon Fibers},'' {\em Journal of the Korean Chemical Society}, vol.~55,
  no.~5, pp.~819--823, 2011.

\bibitem{Govorov2016kinetics}
A.~V. Govorov, A.~A. Galiguzov, N.~A. Tikhonov, A.~Malakho, and A.~D. Rogozin,
  ``{Study of Different Types of Carbon Fiber Oxidation Kinetics},'' {\em
  Refractories and Industrial Ceramics}, vol.~56, no.~6, pp.~605--609, 2016.

\bibitem{ChavezGomez2022flamechemistry}
P.~Ch\'{a}vez-G\'{o}mez, T.~Pelzmann, J.~Zahlawi, L.~{Laberge Lebel}, and
  E.~Robert, ``{Carbon fiber oxidation in combustion environments--Effect of
  flame chemistry and load on bundle failure},'' {\em Materials Today
  Communications}, vol.~31, p.~103560, 2022.

\bibitem{Tong2011kineticstensile}
Y.~Tong, X.~Wang, H.~Su, and L.~Xu, ``{Oxidation kinetics of
  polyacrylonitrile-based carbon fibers in air and the effect on their tensile
  properties},'' {\em Corrosion Science}, vol.~53, no.~8, pp.~2484--2488, 2011.

\bibitem{Feih2012tensilefire}
S.~Feih and A.~P. Mouritz, ``{Tensile properties of carbon fibres and carbon
  fibre-polymer composites in fire},'' {\em {Composites Part A: Applied Science
  and Manufacturing}}, vol.~43, no.~5, pp.~765--772, 2012.

\bibitem{Bertran2013oxidation}
X.~Bertran, C.~Labrugère, M.~A. Dourges, and F.~Rebillat, ``{Oxidation
  Behavior of PAN-based Carbon Fibers and the Effect on Mechanical
  Properties},'' {\em Oxidation of Metals}, vol.~80, no.~3, pp.~299--309, 2013.

\bibitem{Kachold2016preheating}
F.~S. Kachold, R.~Kozera, R.~F. Singer, and A.~Boczkowska, ``{Mechanical
  Properties, Surface Structure, and Morphology of Carbon Fibers Pre-heated for
  Liquid Aluminum Infiltration},'' {\em Journal of Materials Engineering and
  Performance}, vol.~25, no.~4, pp.~1502--1507, 2016.

\bibitem{Vautard2014defect}
F.~Vautard, J.~Dentzer, M.~Nardin, J.~Schultz, and B.~Defoort, ``{Influence of
  surface defects on the tensile strength of carbon fibers},'' {\em Applied
  Surface Science}, vol.~322, pp.~185--193, 2014.

\bibitem{Pickup1986fracture}
I.~M. Pickup, B.~McEnaney, and R.~G. Cooke, ``{Fracture processes in graphite
  and the effects of oxidation},'' {\em Carbon}, vol.~24, no.~5, pp.~535--543,
  1986.

\bibitem{Baker1979insitucatlayst}
R.~T.~K. Baker, ``{In Situ Electron Microscopy Studies of Catalyst Particle
  Behavior},'' {\em Catalysis Reviews—Science and Engineering}, vol.~19,
  no.~2, pp.~161--209, 1979.

\bibitem{Baker1982TammannTemp}
R.~T.~K. Baker, ``{The Relationship between Particle Motion on a Graphite
  Surface and Tammann Temperature},'' {\em Journal of Catalysis}, vol.~78,
  no.~2, pp.~473--476, 1982.

\bibitem{Baker1986catalyst}
R.~T.~K. Baker, ``{Factors controlling the mode by which a catalyst operates in
  the graphite-oxygen reaction},'' {\em Carbon}, vol.~24, no.~6, pp.~715--717,
  1986.

\bibitem{SousaLobo2016kinetics}
L.~{Sousa Lobo} and S.~A.~C. Carabineiro, ``{Kinetics and mechanism of
  catalytic carbon gasification},'' {\em Fuel}, vol.~183, pp.~457--469, 2016.

\bibitem{McMahon1978oxidativeresistance}
P.~McMahon, ``{Oxidative Resistance of Carbon Fibers and their Composites},''
  in {\em Advanced Composite Materials—Environmental Effects} (J.~Vinson,
  ed.), ASTM International, 1978.

\bibitem{Gibbs1979stability}
H.~H. Gibbs, R.~C. Wendt, and F.~C. Wilson, ``{Carbon fiber structure and
  stability studies},'' {\em Polymer Engineering and Science}, vol.~19, no.~5,
  pp.~342--349, 1979.

\bibitem{Eckstein1981oxidation}
B.~H. Eckstein, ``{The oxidation of carbon fibres in air between 230° and
  375° C},'' {\em Fibre Science and Technology}, vol.~14, no.~2, pp.~139--156,
  1981.

\bibitem{Scola1988oxidation}
D.~A. Scola and B.~L. Laube, ``{A Comparison of the Thermo-Oxidative Stability
  of Commercial Graphite Fibers for Composite Applications},'' in {\em {SAE
  Transactions}}, SAE International, 1988.
\newblock {Technical Paper 880111}.

\bibitem{Jones1991boroninfluence}
L.~E. Jones and P.~A. Thrower, ``{Influence of boron on carbon fiber
  microstructure, physical properties, and oxidation behavior},'' {\em Carbon},
  vol.~29, no.~2, pp.~251--269, 1991.

\bibitem{Wu2005boroninhibition}
X.~Wu and L.~R. Radovic, ``{Inhibition of catalytic oxidation of carbon/carbon
  composites by boron-doping},'' {\em Carbon}, vol.~43, no.~8, pp.~1768--1777,
  2005.

\bibitem{Otto1979sulfureffect}
K.~Otto, L.~Bartosiewicz, and M.~Shelef, ``{Catalytic steam gasification of
  graphite: effects of calcium, strontium, and barium with and without
  sulfur},'' {\em Carbon}, vol.~17, no.~4, pp.~351--357, 1979.

\bibitem{Wu2006phosphorusinhibition}
X.~Wu and L.~R. Radovic, ``{Inhibition of catalytic oxidation of carbon/carbon
  composites by phosphorus},'' {\em Carbon}, vol.~44, no.~1, pp.~141--151,
  2006.

\bibitem{Yin1994oxidation}
Y.~Yin, J.~G.~P. Binner, T.~E. Cross, and S.~J. Marshall, ``{The oxidation
  behaviour of carbon fibres},'' {\em Journal of Materials Science}, vol.~29,
  no.~8, pp.~2250--2254, 1994.

\bibitem{Ko1992activation}
T.-H. Ko, P.~Chiranairadul, C.-K. Lu, and C.-H. Lin, ``{The effects of
  activation by carbon dioxide on the mechanical properties and structure of
  PAN-based activated carbon fibers},'' {\em Carbon}, vol.~30, no.~4,
  pp.~647--655, 1992.

\bibitem{Fuertes1996gasification}
A.~B. Fuertes, G.~Marban, and J.~Muñiz, ``{Modelling the gasification of
  carbon fibres},'' {\em Carbon}, vol.~34, no.~2, pp.~223--230, 1996.

\bibitem{Marshall1991topography}
P.~Marshall and J.~Price, ``{Topography of carbon fibre surfaces},'' {\em
  Composites}, vol.~22, no.~5, pp.~388--393, 1991.

\bibitem{Hoffman1992scanning}
W.~P. Hoffman, ``{Scanning probe microscopy of carbon fiber surfaces},'' {\em
  Carbon}, vol.~30, no.~3, pp.~315--331, 1992.

\bibitem{PittmanJr1997nitricacid}
C.~U. Pittman~Jr., G.-R. He, B.~Wu, and S.~D. Gardner, ``{Chemical modification
  of carbon fiber surfaces by nitric acid oxidation followed by reaction with
  tetraethylenepentamine},'' {\em Carbon}, vol.~35, no.~3, pp.~317--331, 1997.

\bibitem{Serp1997vaporgrownCF}
P.~Serp and J.~L. Figueiredo, ``{An investigation of vapor-grown carbon fiber
  behavior towards air oxidation},'' {\em Carbon}, vol.~35, no.~5,
  pp.~675--683, 1997.

\bibitem{Cho1997protection}
D.~Cho, B.~I. Yoon, H.~S. Ha, and Y.~S. Lim, ``{Microscopic Behavior on the
  Protection of Polyacrylonitrile-based Carbon Fibers from Thermal
  Oxidation},'' {\em Polymer Journal}, vol.~29, no.~12, pp.~959--963, 1997.

\bibitem{Tzeng2010resistance}
S.-S. Tzeng, T.-Y. Wu, T.-Y. Chang, C.-T. Yang, C.-L. Chou, and C.-J. Lin,
  ``{Study of Oxidation of Carbon Fibers Using Resistance Measurement},'' {\em
  Journal of Materials Engineering and Performance}, vol.~19, no.~9,
  pp.~1352--1356, 2010.

\bibitem{Iacocca1993catalytic}
R.~G. Iacocca and D.~J. Duquette, ``{The catalytic effect of platinum on the
  oxidation of carbon fibres},'' {\em Journal of Materials Science}, vol.~28,
  no.~4, pp.~1113--1119, 1993.

\bibitem{Panerai2013flow}
F.~Panerai, A.~Martin, N.~N. Mansour, S.~Sepka, and J.~Lachaud, ``{Flow-tube
  Oxidation Experiments on the Carbon Preform of PICA},'' in {\em {$44^{th}$
  AIAA Thermophysics Conference, San Diego, CA, June 24--27, 2013}}, American
  Institute of Aeronautics and Astronautics, 2013.

\bibitem{Panerai2019experimentaloxidation}
F.~Panerai, T.~Cochell, A.~Martin, and J.~D. White, ``{Experimental
  measurements of the high-temperature oxidation of carbon fibers},'' {\em
  International Journal of Heat and Mass Transfer}, vol.~136, pp.~972--986,
  2019.

\bibitem{Cochell2021nanoscaleoxidation}
T.~J. Cochell, R.~R. Unocic, J.~Graña-Otero, and A.~Martin, ``{Nanoscale
  oxidation behavior of carbon fibers revealed with in situ gas cell STEM},''
  {\em Scripta Materialia}, vol.~199, p.~113820, 2021.

\bibitem{ChavezGomez2019fiberoxidation}
P.~{Chávez Gómez}, T.~Pelzmann, E.~Robert, and L.~{Laberge Lebel}, ``{Fuel
  Effect on the Tensile Strength Evolution of Carbon Fibres under Direct Flame
  Attack},'' in {\em {22\textsuperscript{nd} International Conference on
  Composite Materials (ICCM22), Melbourne, Australia, August 11--16}}, 2019.

\bibitem{Cho1993microscopicablation}
D.~Cho, J.~Y. Lee, and B.~I. Yoon, ``{Microscopic observations of the ablation
  behaviours of carbon fibre/phenolic composites},'' {\em Journal of Materials
  Science Letters}, vol.~12, no.~24, pp.~1894--1896, 1993.

\bibitem{Eibl2017respirablefibers}
S.~Eibl, ``{Potential for the formation of respirable fibers in carbon fiber
  reinforced plastic materials after combustion},'' {\em Fire and Materials},
  vol.~41, no.~7, pp.~808--816, 2017.

\bibitem{Bell1979fiberburning}
V.~L. Bell, ``{Release of carbon fibers from burning composites},'' in {\em
  {Assessment of Carbon Fiber Electrical Effects}}, pp.~29--57, National
  Aeronautics and Space Administration, December 1979.
\newblock {NASA Conference Publication 2119}.

\bibitem{AC20-135ch1}
{Federal Aviation Administration}, ``{Powerplant Installation and Propulsion
  System Component Fire Protection Test Methods, Standards and Criteria with
  change 1},'' 10 2018.
\newblock {Advisory Circular AC 20-135 Ch. 1}.

\bibitem{ISO2685}
{International Organization for Standardization}, ``{Aircraft--Environmental
  test procedure for airborne equipment--Resistance to fire in designated fire
  zones},'' December 1998.
\newblock {ISO 2685:1998(E)}.

\bibitem{FAAFireHandbook}
{Federal Aviation Administration}, ``{Aircraft Materials Fire Test Handbook}.''
  \url{https://www.fire.tc.faa.gov/Handbook}.
\newblock {[Online; accessed April 18, 2022]}.

\bibitem{Hughes1962topography}
E.~G. Hughes and J.~M. Thomas, ``{Topography of oxidized graphite crystals},''
  {\em Nature}, vol.~193, no.~4818, pp.~838--840, 1962.

\bibitem{Thomas1964localizedoxidation}
J.~M. Thomas and E.~E.~G. Hughes, ``{Localized oxidation rates on graphite
  surfaces by optical microscopy},'' {\em Carbon}, vol.~1, no.~2, pp.~209--214,
  1964.

\bibitem{Chang1990monolayer}
H.~Chang and A.~J. Bard, ``{Formation of Monolayer Pits of Controlled Nanometer
  Size on Highly Oriented Pyrolytic Graphite by Gasification Reactions as
  Studied by Scanning Tunneling Microscopy},'' {\em Journal of the American
  Chemical Society}, vol.~112, no.~11, pp.~4598--4599, 1990.

\bibitem{Chang1991scanning}
H.~Chang and A.~J. Bard, ``{Scanning Tunneling Microscopy Studies of
  Carbon--Oxygen Reactions on Highly Oriented Pyrolytic Graphite},'' {\em
  Journal of the American Chemical Society}, vol.~113, no.~15, pp.~5588--5596,
  1991.

\bibitem{Chu1991gasificationSTM}
X.~Chu and L.~D. Schmidt, ``{Gasification of graphite studied by scanning
  tunneling microscopy},'' {\em Carbon}, vol.~29, no.~8, pp.~1251--1255, 1991.

\bibitem{Stevens1998kinetics}
F.~Stevens, L.~A. Kolodny, and T.~P. {Beebe Jr.}, ``{Kinetics of Graphite
  Oxidation: Monolayer and Multilayer Etch Pits in HOPG Studied by STM},'' {\em
  The Journal of Physical Chemistry B}, vol.~102, no.~52, pp.~10799--10804,
  1998.

\bibitem{Nicholson2003temperature}
K.~T. Nicholson, T.~K. Minton, and S.~J. Sibener, ``{Temperature-dependent
  morphological evolution of HOPG graphite upon exposure to hyperthermal
  O($^3$P) atoms},'' {\em Progress in Organic Coatings}, vol.~47, no.~3-4,
  pp.~443--447, 2003.

\bibitem{Nicholson2005etching}
K.~T. Nicholson, T.~K. Minton, and S.~J. Sibener, ``{Spatially Anisotropic
  Etching of Graphite by Hyperthermal Atomic Oxygen},'' {\em The Journal of
  Physical Chemistry B}, vol.~109, no.~17, pp.~8476--8480, 2005.

\bibitem{Delehouze2011transition}
A.~Delehouzé, F.~Rebillat, P.~Weisbecker, J.-M. Leyssale, J.-F. Epherre,
  C.~Labrugère, and G.~L. Vignoles, ``{Temperature induced transition from
  hexagonal to circular pits in graphite oxidation by O$_2$},'' {\em Applied
  Physics Letters}, vol.~99, no.~4, p.~044102, 2011.

\bibitem{Dobrik2013etching}
G.~Dobrik, L.~Tapasztó, and L.~P. Biró, ``{Selective etching of armchair
  edges in graphite},'' {\em Carbon}, vol.~56, pp.~332--338, 2013.

\bibitem{Shimada2004pointsonsetgasification}
T.~Shimada, H.~Yanase, K.~Morishita, J.-i. Hayashi, and T.~Chiba, ``{Points of
  onset of gasification in a multi-walled carbon nanotube having an imperfect
  structure},'' {\em Carbon}, vol.~42, no.~8-9, pp.~1635--1639, 2004.

\bibitem{Johanek2016realtime}
V.~Johánek, G.~W. Cushing, J.~K. Navin, and I.~Harrison, ``{Real-time
  observation of graphene oxidation on Pt(111) by low-energy electron
  microscopy},'' {\em Surface Science}, vol.~644, pp.~165--169, 2016.

\bibitem{Tomita1974hydrogenation}
A.~Tomita and Y.~Tamai, ``An optical microscopic study on the catalytic
  hydrogenation of graphite,'' {\em The Journal of Physical Chemistry},
  vol.~78, no.~22, pp.~2254--2258, 1974.

\bibitem{Murray2018dynamics}
V.~J. Murray, E.~J. {Smoll Jr.}, and T.~K. Minton, ``{Dynamics of Graphite
  Oxidation at High Temperature},'' {\em The Journal of Physical Chemistry C},
  vol.~122, no.~12, pp.~6602--6617, 2018.

\bibitem{McCarroll1970interaction}
B.~McCarroll and D.~W. McKee, ``{Interaction of atomic hydrogen and nitrogen
  with graphite surfaces},'' {\em Nature}, vol.~225, no.~5234, pp.~722--723,
  1970.

\bibitem{Fu2022pitting}
R.~Fu, S.~Schmitt, and A.~Martin, ``{Thermo-Chemical-Structural Modeling of
  Carbon Fiber Pitting and Failure Mechanism},'' in {\em {AIAA SciTech Forum,
  San Diego, CA \& Virtual, January 3--7}}, American Institute of Aeronautics
  and Astronautics, 2022.

\bibitem{Hexcel_AS4_TDS}
Hexcel Corp., {\em {HexTow\textsuperscript{\textregistered} AS4 Carbon Fiber
  Product Data Sheet}}.
\newblock {CTA 311 JA20}.

\bibitem{Hexcel_IM7_TDS}
Hexcel Corp., {\em {HexTow\textsuperscript{\textregistered} IM7 Carbon Fiber
  Product Data Sheet}}.
\newblock {CTA 351 NV20}.

\bibitem{Hexcel_HM63_TDS}
Hexcel Corp., {\em {HexTow\textsuperscript{\textregistered} HM63 Carbon Fiber
  Product Data Sheet}}.
\newblock {CTA 355 NV20}.

\bibitem{Schneider2012imagej}
C.~A. Schneider, W.~S. Rasband, and K.~W. Eliceiri, ``{NIH Image to ImageJ: 25
  years of image analysis},'' {\em Nature methods}, vol.~9, no.~7,
  pp.~671--675, 2012.

\bibitem{Townes1985slowpoke}
B.~M. Townes and J.~W. Hilborn, ``{The SLOWPOKE-2 reactor with low enrichment
  uranium oxide fuel},'' tech. rep., Atomic Energy of Canada Ltd., 1985.

\bibitem{Abdollahineisiani2018NAA}
M.~{Abdollahi Neisiani}, M.~Latifi, J.~Chaouki, and C.~Chilian, ``{Novel
  approach in $k_0$-NAA for highly concentrated REE Samples},'' {\em Talanta},
  vol.~180, pp.~403--409, 2018.

\bibitem{Morishita1997CNTgasification}
K.~Morishita and T.~Takarada, ``{Gasification behavior of carbon nanotubes},''
  {\em Carbon}, vol.~35, no.~7, pp.~977--981, 1997.

\bibitem{Morishita1999SEMpurificationCNT}
K.~Morishita and T.~Takarada, ``{Scanning electron microscope observation of
  the purification behaviour of carbon nanotubes},'' {\em Journal of Materials
  Science}, vol.~34, no.~6, pp.~1169--1174, 1999.

\bibitem{Weigand2003}
P.~Weigand, R.~Lückerath, and W.~Meier, ``{Documentation of flat premixed
  laminar CH4/air standard flames: Temperatures and species concentrations},''
  2003.
\newblock {Retrieved on 7-May-2020}.

\bibitem{Boehm2012freeradicals}
H.~P. Boehm, ``{Free radicals and graphite},'' {\em Carbon}, vol.~50, no.~9,
  pp.~3154--3157, 2012.

\bibitem{Arnold2019catalysts}
R.~A. Arnold and J.~M. Hill, ``{Catalysts for gasification: a review},'' {\em
  Sustainable Energy \& Fuels}, vol.~3, no.~3, pp.~656--672, 2019.

\bibitem{Amariglio1966combustioncatalytique}
H.~Amariglio and X.~Duval, ``{Étude de la combustion catalytique du
  graphite},'' {\em Carbon}, vol.~4, no.~3, pp.~323--332, 1966.

\bibitem{Lieberman1972impurities}
M.~L. Lieberman and G.~T. Noles, ``{Impurity effects in carbon fibres},'' {\em
  Journal of Materials Science}, vol.~7, no.~6, pp.~654--662, 1972.

\bibitem{PatentUS4551487vanadium}
C.~A. Gaulin and H.~A. Katzman, ``{Enhanced carbon fiber combustion using a
  catalyst},'' 11 1985.
\newblock US Patent 4,551,487.

\bibitem{Rosner1968oxidation}
D.~E. Rosner and H.~D. Allendorf, ``{Comparative Studies of the Attack of
  Pyrolytic and Isotropic Graphite by Atomic and Molecular Oxygen at High
  Temperatures},'' {\em AIAA Journal}, vol.~6, no.~4, pp.~650--654, 1968.

\bibitem{Neeft1998catalytic}
J.~P.~A. Neeft, M.~Makkee, and J.~A. Moulijn, ``{Catalytic oxidation of carbon
  black-—I. Activity of catalysts and classification of oxidation
  profiles},'' {\em Fuel}, vol.~77, no.~3, pp.~111--119, 1998.

\bibitem{Mckee1975catalytic}
D.~W. McKee and D.~Chatterji, ``{The catalytic behavior of alkali metal
  carbonates and oxides in graphite oxidation reactions},'' {\em Carbon},
  vol.~13, no.~5, pp.~381--390, 1975.

\bibitem{Bevilacqua2015CCbrakes}
M.~Bevilacqua, A.~Babutskyi, and A.~Chrysanthou, ``{A review of the catalytic
  oxidation of carbon--carbon composite aircraft brakes},'' {\em Carbon},
  vol.~95, pp.~861--869, 2015.

\bibitem{Yoshida1990exfoliation}
A.~Yoshida, Y.~Hishiyama, and M.~Inagaki, ``{Exfoliation of vapor-grown
  graphite fibers as studied by scanning electron microscope},'' {\em Carbon},
  vol.~28, no.~4, pp.~539--543, 1990.

\bibitem{Takaku1990structure}
A.~Takaku and M.~Shioya, ``{X-ray measurements and the structure of
  polyacrylonitrile- and pitch-based carbon fibres},'' {\em Journal of
  Materials Science}, vol.~25, no.~11, pp.~4873--4879, 1990.

\bibitem{Rouquerol1994IUPACpores}
J.~Rouquerol, D.~Avnir, C.~W. Fairbridge, D.~H. Everett, J.~M. Haynes,
  N.~Pernicone, J.~D.~F. Ramsay, K.~S.~W. Sing, and K.~K. Unger,
  ``{Recommendations for the characterization of porous solids (Technical
  Report)},'' {\em Pure and Applied Chemistry}, vol.~66, no.~8, pp.~1739--1758,
  1994.

\bibitem{Salatino1993percolative}
P.~Salatino, F.~Miccio, and L.~Massimilla, ``{Combustion and Percolative
  Fragmentation of Carbons},'' {\em Combustion and Flame}, vol.~95, no.~4,
  pp.~342--350, 1993.

\bibitem{Blyholder1958kinetic}
G.~Blyholder, J.~S. {Binford, Jr.}, and H.~Eyring, ``{A kinetic theory for the
  oxidation of carbonized filaments},'' {\em The Journal of Physical
  Chemistry}, vol.~62, no.~3, pp.~263--267, 1958.

\bibitem{Rybak1995oxidation}
W.~Rybak, P.~Chambrion, and J.~Lahaye, ``{Oxidation of carbon black particles
  in a premixed flame under pressure},'' {\em Carbon}, vol.~33, no.~3,
  pp.~259--264, 1995.

\end{thebibliography}

\end{document}

\maketitle

\begin{center}
\fbox{\Large --SUPPLEMENTARY MATERIAL--} 
\end{center}
%\begin{document}

% \begin{center}
% {\large --SUPPLEMENTARY MATERIAL--} 
% \\[0.5\baselineskip]
% {\large\bfseries Carbon fiber damage evolution under flame attack and the role of impurities} 
% \\[1.5\baselineskip]
% {by}
% \\[0.25\baselineskip]
% {Pablo Chávez-Gómez, Tanja Pelzmann, Darren R. Hall, Cornelia Chilian, Louis Laberge Lebel and Étienne Robert}
% \end{center}

%\vspace{2\baselineskip}
\clearpage

\section{Elemental analysis: NAA results}

\begin{figure}[!h]
    \centering
    \includegraphics[width=0.925\columnwidth,height=0.90\textheight,keepaspectratio]{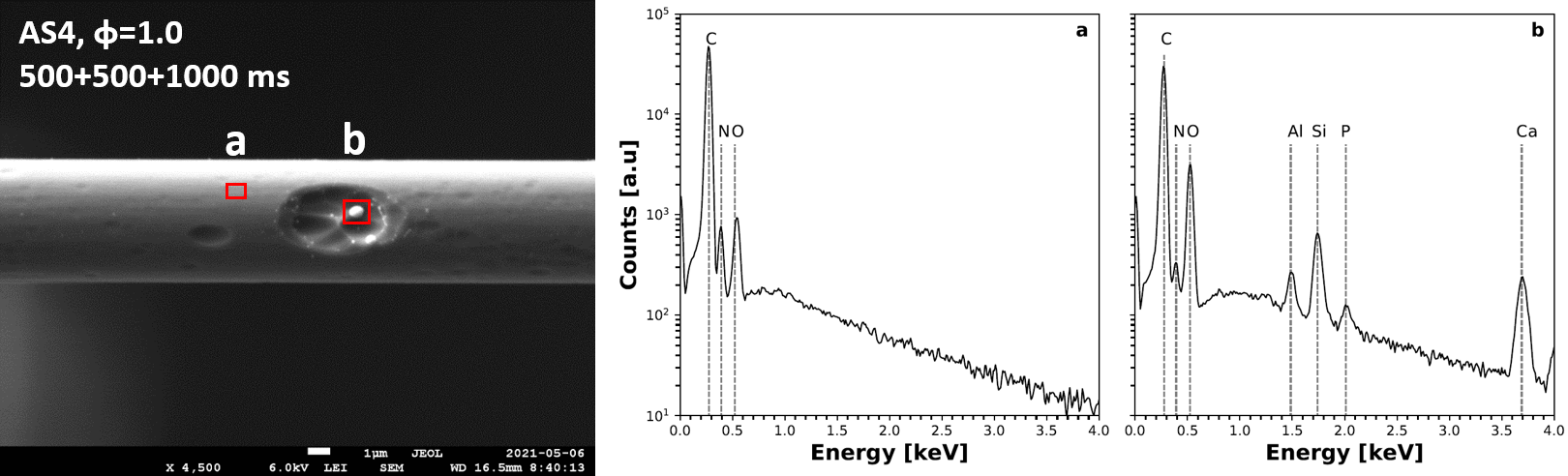}
    \caption{AS4 fiber's SEM image (left) with EDS spectra from: a) smooth skin region and b) mega pit with residue and internal sub-pits.}
    \label{fig:SuppMatl_EDS1_AS4_Pits}
\end{figure}

\begin{figure}[!h]
    \centering
    \includegraphics[width=0.925\columnwidth,height=0.90\textheight,keepaspectratio]{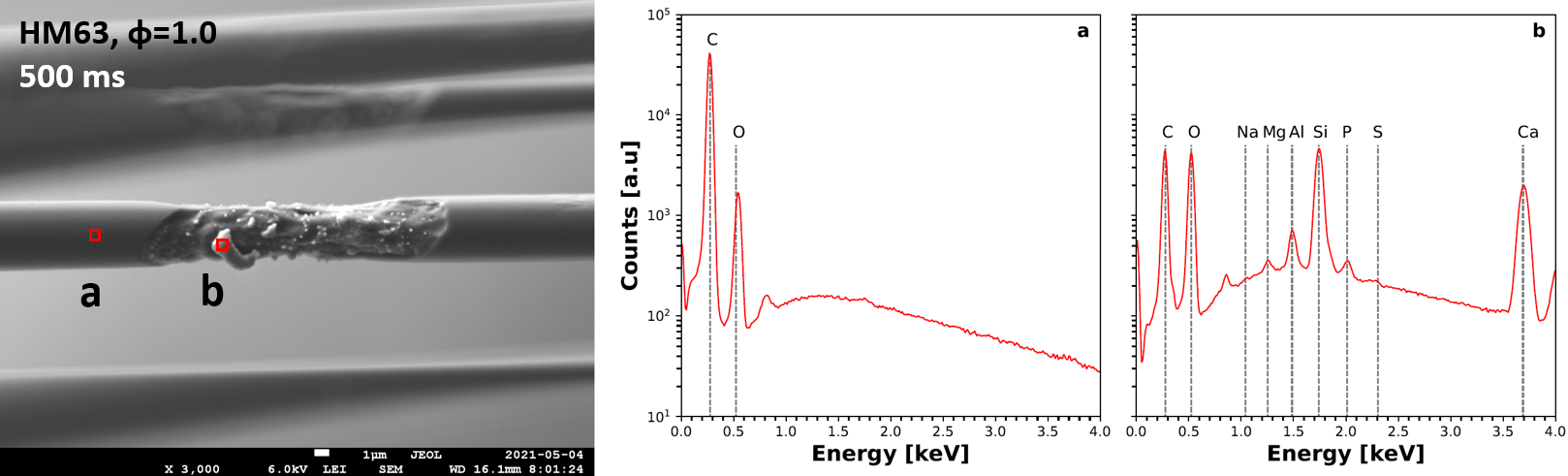}
    \caption{HM63 fiber's SEM image (left) with EDS spectra from: a) smooth skin region and b) Si-based flake within the hourglass-like damaged region.}
    \label{fig:SuppMatl_EDS3_HM63_Hourglass_Damage}
\end{figure}

Table \ref{tbl:NAA_Impurities} shows the impurity concentrations determined by Neutron Activation Analysis (NAA) for each fiber type. Values shown as XXX (XXX) denote concentration and uncertainty of confirmed elements. Values preceded by $<$ correspond to elements that were not possible to be confirmed, although their presence is possible at lower concentrations. For instance, silicon (Si) concentration was not possible to be determined. It could represent the second impurity in terms of prevalence for the three fibers. Its exact concentrations remains undetermined although Figs. \ref{fig:SuppMatl_EDS1_AS4_Pits}b  \ref{fig:SuppMatl_EDS3_HM63_Hourglass_Damage}b reveal its presence in amorphous residues over damaged areas, supporting the main article's observations.

\begin{table}[h]
\centering
\caption{Fiber impurity levels obtained by Neutron Activation Analysis (NAA).} 
\label{tbl:NAA_Impurities}
\resizebox*{!}{0.72\textheight}{
\begin{tabular}{cccc}
\hline
\multirow{2}{*}{Element} & \multicolumn{3}{c}{Concentration {[}ppm{]}} \\
 & AS4 & IM7 & HM63 \\ \hline
Na & 1024 (41) & 1079 (43) & 11.9 (0.5) \\
Mg & 2.74 (0.75) & 3.19 (0.74) & 2.70 (0.24) \\
Al & 9.20 (0.40) & 12.2 (0.5) & 5.63 (0.24) \\
Si & \textless~420 & \textless~450 & \textless~120 \\
S & \textless~320 & \textless~360 & \textless~60 \\
Cl & 4.72 (0.34) & 6.57 (0.42) & 18.6 (0.8) \\
K & \textless~70 & \textless~40 & \textless~20 \\
Ca & 16.2 (2.2) & 14.1 (2.6) & 20.1 (1.3) \\
Sc & \textless~0.007 & \textless~0.008 & \textless~0.003 \\
Ti & \textless~2 & \textless~1 & 0.698 (0.107) \\
V & \textless~0.007 & \textless~0.008 & 0.103 (0.004) \\
Cr & \textless~0.5 & \textless~0.9 & \textless~0.4 \\
Mn & 0.0447 (0.0029) & 0.0713 (0.0200) & 0.0381 (0.0040)\\
Fe & \textless~70 & \textless~70 & \textless~20 \\
Co & \textless~0.2 & \textless~0.2 & \textless~0.07 \\
Ni & \textless~20 & \textless~8 & \textless~20 \\
Cu & \textless~0.5 & \textless~0.9 & 0.150 (0.039) \\
Zn & \textless~8 & \textless~7 & \textless~2 \\
As & \textless~0.02 & \textless~0.02 & \textless~0.006 \\
Se & \textless~0.4 & \textless~0.4 & \textless~0.4 \\
Br & 0.0815 (0.0166) & 0.180 (0.019) & 0.064 (0.005) \\
Rb & \textless~2 & \textless~2 & \textless~0.4 \\
Zr & \textless~50 & \textless~30 & \textless~20 \\
Mo & \textless~0.2 & \textless~0.09 & \textless~0.08 \\
Ag & \textless~0.3 & \textless~0.4 & \textless~0.09 \\
Cd & \textless~0.2 & \textless~0.2 & \textless~0.09 \\
In & \textless~0.001 & \textless~0.0008 & \textless~0.0002 \\
Sn & \textless~0.7 & \textless~0.5 & \textless~0.2 \\
Sb & 0.0126 (0.0019) & 0.0361 (0.0023) & 0.0089 (0.00088) \\
I & 0.0440 (0.0056) & 0.0781 (0.0072) & 1.54 (0.06) \\
Cs & \textless~0.08 & \textless~0.06 & \textless~0.02 \\
Ba & \textless~6 & \textless~5 & \textless~2 \\
La & \textless~0.008 & \textless~0.01 & \textless~0.006 \\
Hf & \textless~0.04 & \textless~0.04 & \textless~0.02 \\
W & \textless~0.03 & \textless~0.04 & \textless~0.04 \\
Au & \textless~0.0002 & \textless~0.0003 & \textless~0.0002 \\
Hg & 0.266 (0.055) & 0.113 (0.057) & \textless~0.05 \\
Th & \textless~0.03 & \textless~0.3 & \textless~0.02 \\
U & \textless~0.02 & \textless~0.02 & \textless~0.02 \\\hline
\multicolumn{4}{l}{\footnotesize{Concentration (Uncertainty)}} 
\end{tabular}}
\end{table}

\clearpage

\section{Elemental analysis: EDS spectra}

The following figures show additional EDS spectra of some of the same fibers shown in the main text. Relatively clean and impurity-rich regions yield contrasting elemental signatures.

\begin{figure}[!h]
    \centering
    \includegraphics[width=0.925\columnwidth,height=0.90\textheight,keepaspectratio]{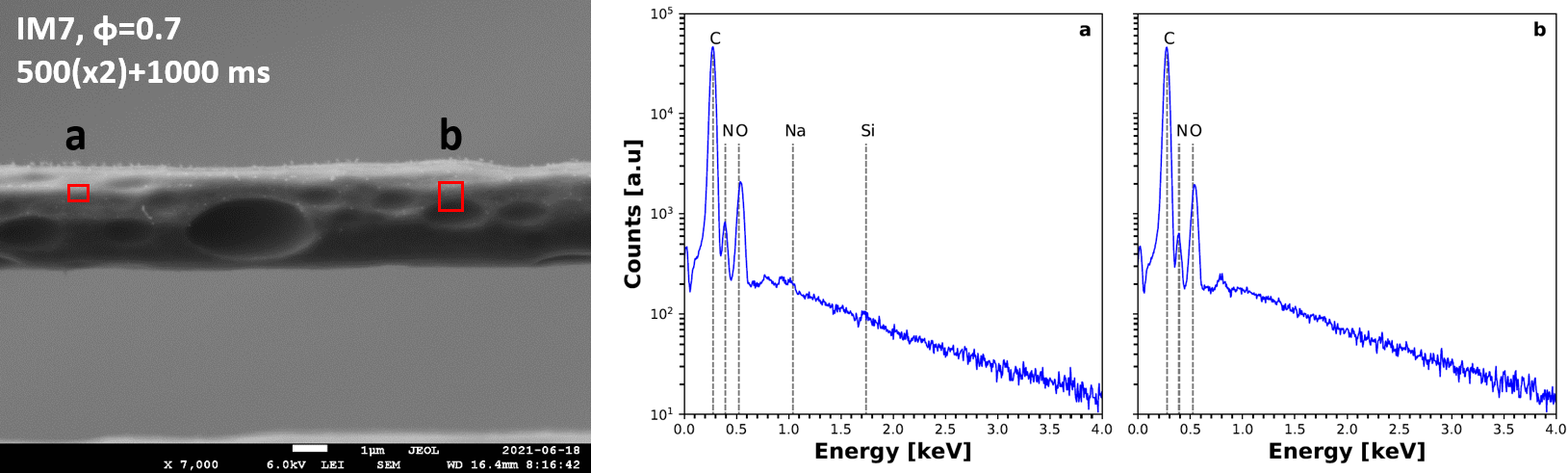}
    \caption{IM7 fiber's SEM image (left) with EDS spectra from: a) smooth fiber portion and b) large pit inner and outer portions.}
    \label{fig:SuppMatl_EDS2_IM7_Pits}
\end{figure}

\begin{figure}[!h]
    \centering
    \includegraphics[width=0.925\columnwidth,height=0.90\textheight,keepaspectratio]{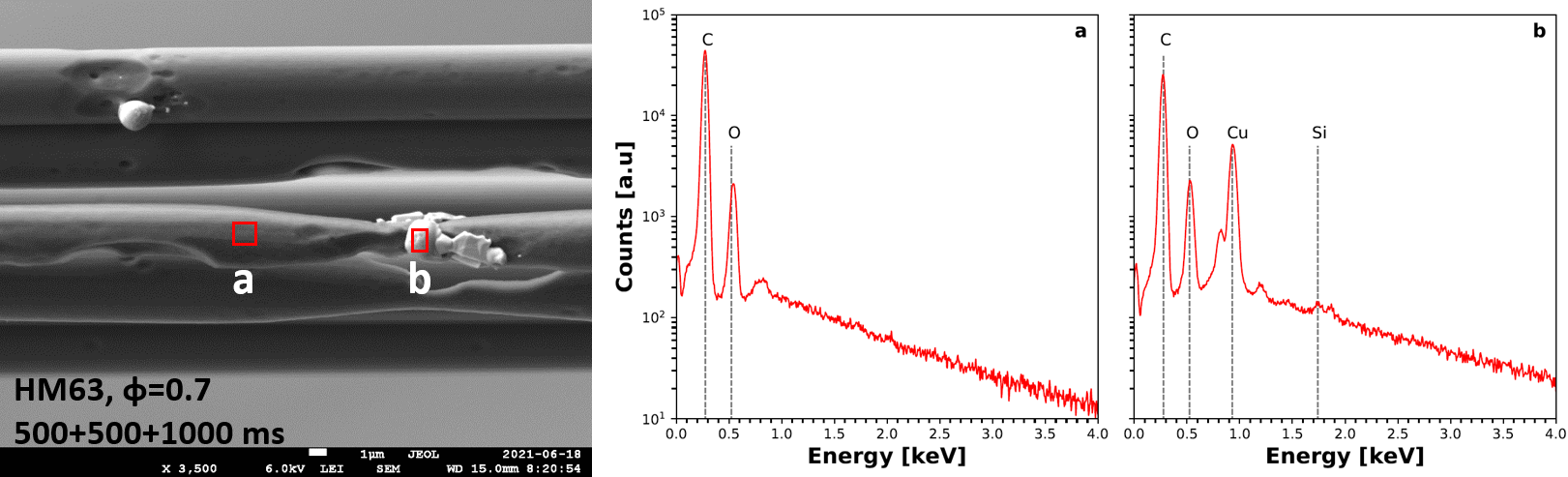}
    \caption{HM63 fiber's SEM image (left) with EDS spectra from: a) smooth skin region and b) amorphous Cu-based impurity within the hourglass-like damaged region.}
    \label{fig:SuppMatl_EDS4_HM63_Cu_Hourglass}
\end{figure}

\begin{figure}[!h]
    \centering
    \includegraphics[width=0.925\columnwidth,height=0.90\textheight,keepaspectratio]{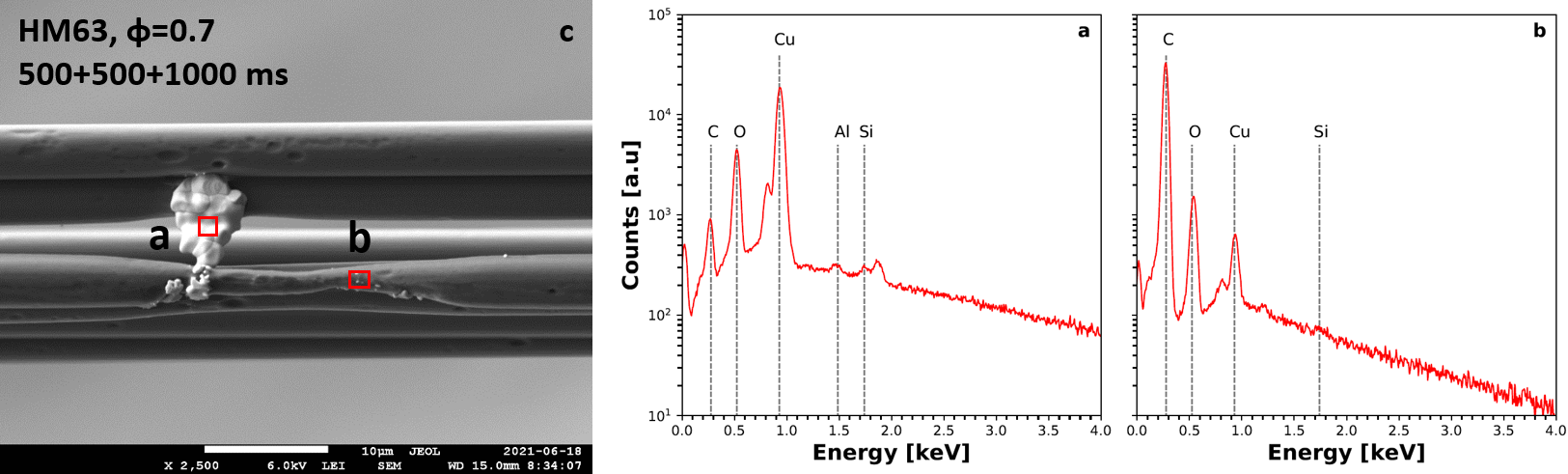}
    \caption{HM63 fiber's SEM image (left) with EDS spectra from: a) amorphous Cu-based impurity seemingly stuck between fibers and b) thinned  region within the hourglass-like damaged region.}
    \label{fig:SuppMatl_EDS5_HM63_Cu_Hourglass_Popcorn}
\end{figure}

\begin{figure}[!h]
    \centering
    \includegraphics[width=0.925\columnwidth,height=0.90\textheight,keepaspectratio]{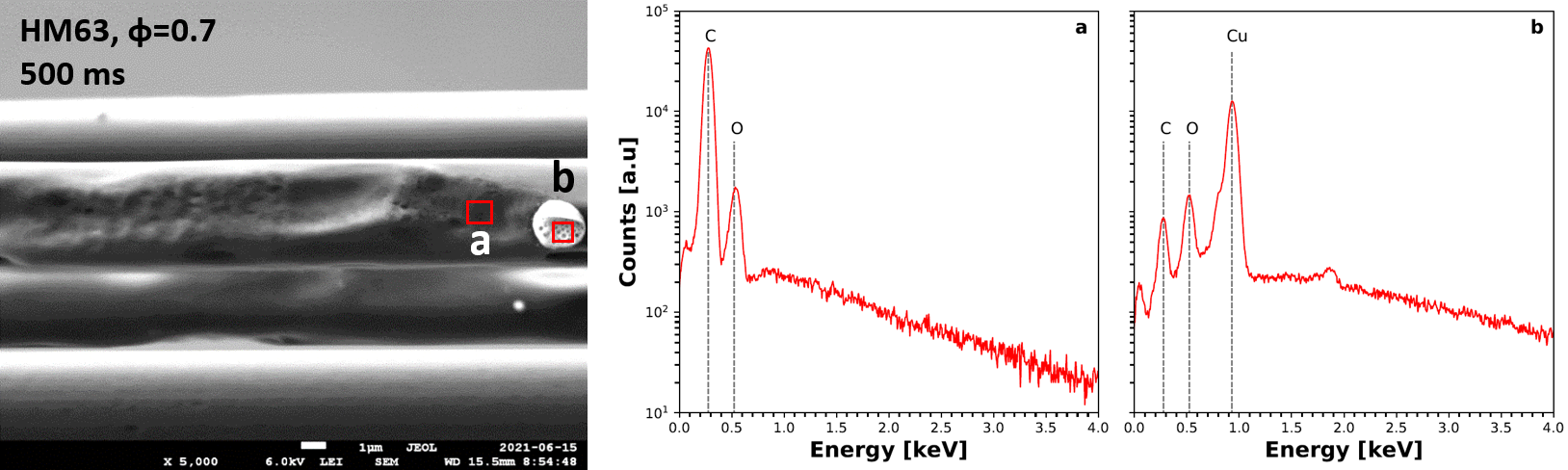}
    \caption{HM63 fiber's SEM image (left) with EDS spectra from: a) damaged region by mobile impurity and b) Cu-based mobile impurity.}
    \label{fig:SuppMatl_EDS6_HM63_Mobile_Porous_Cu_Ball}
\end{figure}